\documentclass[fleqn,11pt,twoside]{article}

\usepackage{amsmath,amsthm,amssymb,color,xcolor,epsfig,graphics,subfigure}

\usepackage{latexsym, lscape}
\usepackage{array,tabularx,booktabs}
\usepackage{longtable}
\usepackage{url}
\usepackage{cancel,xspace}
\usepackage{cite}

\DeclareMathOperator{\Aut}{Aut}
\DeclareMathOperator{\Int}{Int}
\DeclareMathOperator{\ad}{ad}
\DeclareMathOperator{\gl}{\mathfrak{gl}}

\newtheorem{remark}{Remark}
\newtheorem*{algebra}{Algebra}
\newcommand{\smallsize}{\fontsize{8pt}{11pt}\selectfont}
\makeatletter
\newcommand{\copyrightnote}[2]{{\renewcommand{\thefootnote}{}
 \footnotetext{\small\it
\begin{flushleft}
 \copyright \ #1   #2  
\end{flushleft}}}}

\newcommand{\Name}[1]{\begin{flushleft}
                       \LARGE \bf #1
                       \end{flushleft}\vspace{-3mm}}

\newcommand{\Author}[1]{\begin{flushleft}
                       \it #1 \end{flushleft}}

\newcommand{\Address}[1]{\begin{flushleft}
                       \it #1 \end{flushleft}}

\newcommand{\Date}[1]{\begin{flushleft}
                      \small  \it #1 \end{flushleft}}

\newcommand{\evenhead}{}
\newcommand{\oddhead}{}

\renewcommand{\@evenhead}{
\hspace*{-3pt}\raisebox{-15pt}[\headheight][0pt]{\vbox{\hbox to \textwidth
{\thepage \hfil \evenhead}\vskip4pt \hrule}}}
\renewcommand{\@oddhead}{
\hspace*{-3pt}\raisebox{-15pt}[\headheight][0pt]{\vbox{\hbox to \textwidth
{\oddhead \hfil \thepage}\vskip4pt\hrule}}}
\renewcommand{\@evenfoot}{}
\renewcommand{\@oddfoot}{}

\long\def\@makecaption#1#2{%
  \vskip\abovecaptionskip
  \sbox\@tempboxa{\small \textbf{#1.}\ \ #2}%
  \ifdim \wd\@tempboxa >\hsize
    {\small \textbf{#1.}\ \ #2}\par
  \else
    \global \@minipagefalse
    \hb@xt@\hsize{\hfil\box\@tempboxa\hfil}%
  \fi
  \vskip\belowcaptionskip}

\newcommand{\JNMPnumberwithin}[3][\arabic]{%
  \@ifundefined{c@#2}{\@nocounterr{#2}}{%
    \@ifundefined{c@#3}{\@nocnterr{#3}}{%
      \@addtoreset{#2}{#3}%
      \@xp\xdef\csname the#2\endcsname{%
        \@xp\@nx\csname the#3\endcsname .\@nx#1{#2}}}}%
}

\newcommand{\resetfootnoterule} {
  \renewcommand\footnoterule{%
  \kern-3\p@
  \hrule\@width.4\columnwidth
  \kern2.6\p@}
}

\renewcommand{\footnoterule}{}

\newcommand{\symbolie}{\texttt{SymboLie}\xspace}

\makeatother

\theoremstyle{definition}
\newtheorem{definition}{Definition}
\newtheorem{example}{Example} 

\begin{document}

\renewcommand{\evenhead}{ {\LARGE\textcolor{blue!10!black!40!green}{{\sf \ \ \ ]ocnmp[}}}\strut\hfill 
L.~Amata, F.~Oliveri, E.~Sgroi
}
\renewcommand{\oddhead}{ {\LARGE\textcolor{blue!10!black!40!green}{{\sf ]ocnmp[}}}\ \ \ \ \  
Computation of optimal subalgebras of 3D and 4D real Lie algebras
}

\thispagestyle{empty}
\newcommand{\FistPageHead}[3]{
\begin{flushleft}
\raisebox{8mm}[0pt][0pt]
{\footnotesize \sf
\parbox{150mm}{{\textcolor{blue!10!black!40!green}{{\bf Open Communications in Nonlinear Mathematical Physics}}}
\ \ {Special Issue: Bluman}, 2025\\[0.1cm]
\strut\hfill 
ocnmp:16985,
pp #2\hfill {\sc #3}}}\vspace{-13mm}
\end{flushleft}}

\FistPageHead{47}{\pageref{firstpage}-\pageref{lastpage}}{ \ \ }

\strut\hfill

\strut\hfill

\copyrightnote{The authors. Distributed under a Creative Commons Attribution 4.0 International License}

\begin{center}

{\bf {\large A Special OCNMP Issue in Honour of George W Bluman}}
\end{center}

\smallskip

\Name{Symbolic computation of optimal systems of subalgebras
of three- and four-dimensional real Lie algebras}

\Author{Luca Amata, Francesco Oliveri\footnote{Corresponding author.}, Emanuele Sgroi}

\Address{MIFT Department, University of Messina\\
viale F. Stagno d'Alcontres 31, 98166 Messina, Italy\\
email: lamata@unime.it; foliveri@unime.it; emasgroi@unime.it}

\Date{Received November 25, 2025; Accepted December 7, 2025}

\setcounter{equation}{0}

\smallskip

\noindent
{\bf Citation format for this Article:}\newline
L.~Amata, F.~Oliveri, E.~Sgroi,
Symbolic computation of optimal systems of subalgebras
of three- and four-dimensional real Lie algebras,
{\it Open Commun. Nonlinear Math. Phys.}, Special Issue:\,Bluman, ocnmp:16985, \pageref{firstpage}-\pageref{lastpage}, 2025.

\strut\hfill

\noindent
{\bf The permanent Digital Object Identifier (DOI) for this Article:}\newline
{\it 10.46298/ocnmp.16985}
\strut\hfill

\begin{abstract}
\noindent 
The complete optimal systems of subalgebras of all nonisomorphic three- and four-dimensional real Lie algebras are analyzed by the program \symbolie  running in the computer algebra system \emph{Wolfram Mathematica}\texttrademark. The approach uses the  definition of $p$-families of Lie subalgebras whose set  can be partitioned by introducing a binary relation (reflexive and transitive, though not necessarily symmetric) induced by inner automorphisms of the Lie algebra. The results, produced in a few minutes by \symbolie, represent a good test for the program; in fact, except for minor differences that are discussed, the results confirm those given in 1977 in a paper by Patera and Winternitz. 

symbolic computation.

\noindent\textbf{AMS[2020]}: 17B40, 35B06, 68W30.
\end{abstract}

\label{firstpage}


\section{Introduction}\label{sec:intro}
In the late 19th century, Sophus Lie \cite{Lie1888,Lie1891,Lie_translation} studied the invariance of differential equations with respect to continuous groups of transformations, now referred to as Lie groups. A widely used application of Lie groups admitted by partial differential equation consists in the determination of invariant solutions \cite{Ovsiannikov,Olver,OlverRosenau1987,BlumanKumei,Meleshko,Oliveri-symmetry}. 
Different Lie subgroups of symmetries of partial differential equations in principle lead to different invariant solutions; nevertheless, equivalent subgroups (with respect to the conjugacy relation) lead to equivalent invariant solutions \cite{Ovsiannikov,Olver} linked each other by the action of some subgroup. So, it is appropriate to classify invariant solutions,
\emph{i.e.}, to separate solutions into classes of equivalent solutions.
A Lie group of symmetries admitted by a differential equation is associated to a Lie algebra 
\cite{Humphreys,Erdmann,Kirillov,Snobl-Winternitz} of its infinitesimal generators, and Lie subalgebras correspond to Lie subgroups. As a consequence, the problem of identifying inequivalent subgroups of a Lie group can be addressed by classifying inequivalent Lie subalgebras; the latter problem can be more easily approached from an algorithmic point of view \cite{AmataOliveri1,AmataOliveri2,Symbolie}.

In the literature, some methods for finding optimal systems of Lie subalgebras of finite-dimensional Lie algebras have been described. For instance, Ovsiannikov \cite[Ch.~IV, Sects.~7-9]{Ovsiannikov} illustrates how to use the \emph{common inner automorphisms} to obtain the inequivalent subalgebras through some simplifications. Moreover, Olver \cite[Ch.~3, Sect.~3.3]{Olver} describes a method using the sign of the \emph{Killing form} (which is left invariant by the adjoint action) to classify the conjugacy classes. More recently, Olver's approach has been generalized by using other invariants in addition to the Killing form (see, for instance, \cite{HuLiChen}). Another interesting approach can be found in \cite{PWZ} where Patera, Winternitz and Zassenhaus used cohomology theory to obtain the conjugacy classes.

Most of the methods used to address this problem involve choices that may be rather hard to automate algorithmically in a computer algebra system. Some authors \cite{{HuLiChen,ZhangHanChen}} claimed the implementation of computer algebra algorithms for determining optimal systems of Lie subalgebras; nevertheless, these algorithms are not systematic, and the results are obtained using the computer algebra system interactively as a symbolic calculator. Recently, a SageMath \cite{Sagemath} program for classifying Lie subalgebras has been described \cite{Yakhno}.

In \cite{AmataOliveri1, AmataOliveri2, Symbolie}, a new algorithmic approach for determining optimal systems of Lie subalgebras has been proposed. The algorithm has been implemented in the Wolfram Mathematica\texttrademark\ \cite{Wolfram} package \symbolie (freely available at the url \url{https://mat521.unime.it/oliveri}).
The main focus of the present paper, intended to be a meaningful test for the program,  is to analyze all three- and four-dimensional real Lie algebras classified in 1977 by Patera and Winternitz \cite{PW}, and compare the optimal systems obtained by \symbolie with those listed in \cite{PW}.

The paper is organized as follows. In Section~\ref{sec:preliminaries}, after recalling some basic algebraic notions,  we introduce the notion of $p$-family of Lie subalgebras, and define the relation linking the different $p$-families. Section~\ref{sec:symbolie} briefly illustrates how the package has to be used.
In Sections~\ref{sec:3D} and \ref{sec:4D}, we determine the optimal systems of Lie algebras of dimension three and four  \cite{PW} by using the \symbolie package. We discuss in detail the Lie algebras for which the program gives solutions a little bit different from those provided in \cite{PW}. We also collect in two tables the results provided by \symbolie.
Finally, Section~\ref{sec:concl} contains our conclusions and perspectives.

\section{Theoretical preliminaries and notation}
\label{sec:preliminaries}
Let $\mathcal{L}$ be an $r$-dimensional real Lie algebra with basis\footnote{We will denote the elements of the basis of the Lie algebra as $\Xi_i$, in agreement with \symbolie notation.} $\mathcal{B}=\{\Xi_1,\ldots,\Xi_r\}$. For each ordered pair $(\Xi_\alpha,\Xi_\beta)$,  its Lie bracket can be written as a linear combination of elements of $\mathcal{B}$, \emph{i.e.}, there exist constant values $C_{\alpha\beta}^\gamma\in \mathbb{R}$, $\alpha,\beta,\gamma=1,\ldots,r$, such that
\[
[\Xi_\alpha,\Xi_\beta] = \sum_{\gamma=1}^r C_{\alpha\beta}^\gamma \Xi_\gamma,
\]
where the structure constants $C_{\alpha\beta}^\gamma$ determine the Lie algebra completely.

By taking two elements $X,Y\in\mathcal{L}$, they can be decomposed according to the basis in the form
\[
X=\sum_{\alpha=1}^r f^\alpha \Xi_\alpha,\qquad Y=\sum_{\alpha=1}^r g^\alpha \Xi_\alpha,
\]
where $f^\alpha$ and $g^\alpha$ are suitable elements belonging to $\mathbb{R}$, whence
\begin{equation}\label{eq:structure}
\left[X,Y\right]=\sum_{\alpha,\beta=1}^r f^\alpha g^\beta\left[\Xi_\alpha,\Xi_\beta\right]=
\sum_{\alpha,\beta,\gamma=1}^r f^\alpha g^\beta C_{\alpha\beta}^\gamma \Xi_\gamma.
\end{equation}

Relation~\eqref{eq:structure} implies that, for the \emph{coordinates} $\mathbf{f}\equiv(f^1,\ldots,f^r)$ and $\mathbf{g}\equiv(g^1,\ldots,g^r)$ of the elements $X$ and $Y$ with respect to the basis  $\{\Xi_1,\ldots,\Xi_r\}$, we may introduce a Lie bracket in $\mathbb{R}^r$ such that
\[
\left[\mathbf{f},\mathbf{g}\right]^\gamma=\sum_{\alpha,\beta=1}^r f^\alpha g^\beta C_{\alpha\beta}^\gamma, \qquad \gamma=1,\ldots,r.
\]

Given a Lie algebra $\mathcal{L}$, a \emph{Lie subalgebra} is a vector subspace $\mathcal{L}^\prime\subseteq \mathcal{L}$ such that $[X,Y] \in \mathcal{L}^\prime$ for all $X,Y\in \mathcal{L}^\prime$.

Trivially, any $X\in\mathcal{L}$ spans
a one-dimensional subalgebra with elements $aX$ ($a\in \mathbb{R}$), \emph{i.e.}, one-dimensional subalgebras 
always exist in any Lie algebra $\mathcal{L}$. On the contrary, as far as the two-dimensional subalgebras of a real Lie algebra $\mathcal{L}$ are concerned, their existence is not always guaranteed. For instance, the three-dimensional Lie algebra associated to the group $\text{SO}(3)$, characterizing the rotations in a three-dimensional space, does not possess two-dimensional subalgebras.

A Lie subalgebra of dimension $1<d<r$  is spanned by the elements
\begin{equation} \label{sub_s}
X_i=\sum_{\alpha=1}^r h_i^\alpha \Xi_\alpha, \qquad i=1,\ldots,d,
\end{equation}
where the $d\times r$ matrix $\left\| h_i^\alpha\right\|$ has rank $d$, provided that 
\begin{equation}
\label{comm_s}
\left[X_i,X_j \right] = \sum_{k=1}^d \lambda_{ij}^k X_k, \qquad i,j=1,\ldots,d,
\end{equation}
$\lambda_{ij}^k$ being suitable constants.

By using \eqref{sub_s}, relations \eqref{comm_s} yield
\begin{equation}\label{eq:subalgebras}
\sum_{\alpha,\beta=1}^r h_i^\alpha h_j^\beta C_{\alpha\beta}^\gamma = \sum_{k=1}^d \lambda_{ij}^k h_k^\gamma,  \qquad i,j=1,\ldots, d, \gamma=1,\ldots,r,
\end{equation}
clarifying the fact that the determination of a Lie subalgebra requires the solution of a system of quadratic equations. 
In fact, what is needed is to find the entries
$h_i^\alpha$ of a matrix with rank $d$ such that relations~\eqref{eq:subalgebras} hold true for some constants 
$\lambda_{ij}^k$.

A \emph{representation} of a Lie algebra $\mathcal{L}$ is a Lie algebra homomorphism 
$\varphi\colon \mathcal{L} \to \gl(V)$, where $V$ is a vector space and $\gl(V)$ is the \emph{general linear algebra} of all linear maps of $V$ into itself.
A useful representation is the \emph{adjoint representation} $\ad\colon\mathcal{L}\to \mathfrak{gl}(\mathcal{L})$, sending $X$ to $\ad_X$, with $\ad_X(Y) = [X,Y]$, for all $X,Y\in\mathcal{L}$.

An \emph{automorphism} of $\mathcal{L}$ is an isomorphism of $\mathcal{L}$ onto itself, and the set of all automorphisms of $\mathcal{L}$ form a group, denoted as $\Aut(\mathcal{L})$. Its subgroup generated by 
$\{\exp(\ad_X) \colon X\in \mathcal{L}\}$ is called the \emph{group of inner automorphisms} of $\mathcal{L}$, denoted as $\Int(\mathcal{L})$.

Two Lie subalgebras $\mathcal{L}^\prime$ and $\mathcal{L}^{\prime\prime}$ of a Lie algebra $\mathcal{L}$ are said \emph{equivalent} if there exists an inner automorphism $\phi\in \Int(\mathcal{L})$ such that $\phi(\mathcal{L}^\prime) = \mathcal{L}^{\prime\prime}$.

The given definition establishes an equivalence relation on the set of the Lie subalgebras of a Lie algebra $\mathcal{L}$. So, all subalgebras of $\mathcal{L}$ are decomposed into classes of equivalent algebras, that is the set of subalgebras of $\mathcal{L}$ is partitioned into \emph{conjugacy classes}. A set of the representatives of each class is called an \emph{optimal system of subalgebras}; it is denoted by $\Theta_\mathcal{A}(\mathcal{L})$, where $\mathcal{A} = \Int(\mathcal{L})$.

Hence, the optimal system of subalgebras of a Lie algebra $\mathcal{L}$ with inner automorphisms $\mathcal{A}$ is a set of subalgebras such that \cite{Ovsiannikov}:
\begin{enumerate}
\item there are no two elements of this set which can be transformed into each other by inner automorphisms of the Lie algebra $\mathcal{L}$;
\item any subalgebra of the Lie algebra $\mathcal{L}$ can be transformed into one of subalgebras of the set $\Theta_\mathcal{A}(\mathcal{L})$.
\end{enumerate}
This set is unique in the following sense: if $\Theta_\mathcal{A}$ and $\Theta_\mathcal{A}^\prime$ are two optimal systems, they have the same cardinality, and, if $N\in \Theta_\mathcal{A}$ and $N^\prime\in \Theta_\mathcal{A}^\prime$, then there exists an inner automorphism $\phi$ that such $\phi(N) = N^\prime$.

In order to address from an algorithmic point of view the problem of characterizing optimal systems of Lie subalgebras,
let us recall some definitions from \cite{AmataOliveri2}. Such definitions are useful for implementing a program able to perform automatically the determination of the optimal systems of Lie subalgebras of a finite-dimensional real Lie algebra.

As above underlined,  the determination of a Lie subalgebra requires to solve a system of quadratic equations. This task can be 
accomplished algorithmically in a computer algebra system; nevertheless, such a process may have a high computational cost, and, often, the solutions can be obtained only for low dimensions. 

\subsection{One-dimensional Lie subalgebras}
A one-dimensional Lie subalgebra is completely defined by the coefficients of the linear combination of the generators of the basis of $\mathcal{L}$, say
\[
X=f^1\Xi_1+f^2\Xi_2+\cdots+f^r\Xi_r,
\]
and actions of inner automorphisms can be transferred to the \emph{coordinates} $(f^1,f^2,\ldots,f^r)$ of an $r$-dimensional vector; the method largely used in the literature for obtaining optimal systems of one-dimensional Lie subalgebras consists in using inner automorphisms to obtain from the most general $r$-dimensional vector the maximum possible number of zero coordinates, also using the invariants of the inner automorphisms. More in detail, the method takes a tuple $\{f^1,f^2,\ldots,f^r\}$, and, through \emph{judicious} applications of inner automorphisms, simplifies as many coefficients $f^\alpha$ as possible \cite[Example 3.12]{Olver}. This method needs to solve algebraic equations and make suitable choices during the process to distinguish cases; this is
relatively easy only for low dimensional Lie algebras, and the simplicity of the results obtained is not always clear.

To implement an almost automatic program for the ``subalgebra problem'', let us introduce some new \emph{objects}.

\begin{definition}\label{def:binaryruples}
Let $\mathcal{S}_r=\{0,1\}^r \setminus \{0,0,\ldots,0\}$; the cardinality of this set is $2^r-1$.
\end{definition}
\begin{definition}\label{def:family1D}
Let $\mathcal{L}$ be an $r$-dimensional Lie algebra generated by $\{\Xi_1,\Xi_2,$ $\ldots,\Xi_r\}$, and let 
$\mathbf{f}\equiv\{f^1,f^2,\ldots,f^r\}$ be a tuple of $r$ real functions depending on some real variables belonging to the set $\mathcal{P}$.
Then, the \emph{family of one-dimensional subalgebras}
\[
X=f^{1}s_1 \Xi_{1}+f^{2}s_2 \Xi_{2}+\cdots+f^{r}s_r \Xi_{r},
\]
where $\mathbf{s}\equiv(s_1,\ldots,s_r) \in \mathcal{S}_r$, is called a $p$-\emph{family} of one-dimensional Lie subalgebras of $\mathcal{L}$ if the rank of the Jacobian matrix of $(f^{1}s_1,f^{2}s_2,\ldots,f^{r}s_r)$ with respect to the elements of $\mathcal{P}$ is equal to $p=\sum_{i=1}^r s_i$.
\end{definition}

Definition~\ref{def:family1D} means that the decomposition of $X$ in terms of the elements of the basis of the Lie algebra involves $p$ functionally independent coefficients.

\begin{example}\label{exa:family1D}
\index{Lie algebra!family of one-dimensional subalgebras}
Let $\mathcal{L}_4$ be a four-dimensional Lie algebra spanned by $\{\Xi_1,\Xi_2,\Xi_3,$ $\Xi_4\}$, and consider the family of 
one-dimensional subalgebras 
\[
X=f^{1} \Xi_{1}+f^{2} \Xi_{2},
\] 
with $f^1$ and $f^2$ arbitrary real numbers; in this case, it is $\mathcal{P} = \{f^{1}, f^{2}\}$, and we have a 
2-family of one-dimensional Lie subalgebras. Also the family
\[
X^\prime=(f^{1}\cos(t)-f^2\sin(t)) \Xi_{1}+(f^1\sin(t)+f^{2}\cos(t)) \Xi_{2},
\] 
where $\mathcal{P}=\{f^1,f^2,t\}$, is a 2-family of one-dimensional Lie subalgebras which has to be considered indistinguishable from $X$.
On the contrary, the element 
\[
Y=f^{1} \Xi_{1}+f^{2} \Xi_{2}+ f^{3}\Xi_{3},
\] 
where the components $f^1$, $f^2$ and $f^3$ satisfy the relation $f^1 f^2=(f^3)^2$, does not correspond to a 3-family.
Indeed, although the vector $(f^{1},f^{2},f^{3},0)$ possesses three nonzero components, we do not have a $3$-family because the three components are not independent.
\end{example}

The $p$-family of one-dimensional Lie subalgebras of $\mathcal{L}$
\[
X=f^1 s_1 \Xi_1+f^2 s_2\Xi_2+\cdots+f^r s_r\Xi_r,
\]
where $\mathbf{s}\equiv\{s_1,s_2,\ldots,s_r\}\in \mathcal{S}_r$, $p=\sum_{i=1}^r s_i$, in the following will be represented by the tuple
\begin{equation*}
\left(f^1 s_1, f^2 s_2,\ldots,f^r s_r\right)
\end{equation*}
which in turn can be identified by the integer
$\displaystyle c_X = \sum_{k=1}^r s_k 2^{k-1}$.

Then, let us introduce a relation $\mathcal{R}$ between $p$-families of one-dimensional Lie subalgebras. 

\begin{definition}\label{def:relationpfamily1D}
Let $\mathcal{L}$ be an $r$-dimensional Lie algebra generated by $\{\Xi_1,\dots,\Xi_r\}$, and let ${X}\equiv(f^1s_1,\ldots,f^rs_r)$ and ${Y}\equiv(\tilde{f}^1\tilde{s}_1,\ldots,\tilde{f}^r\tilde{s}_r) $ be two different families of one-dimensional Lie subalgebras.
It is
\[
X\,\mathcal{R}\,Y\quad \text{or} \quad (X,Y)\in\mathcal{R}
\]
if the action of an inner automorphism on $X$ produces a family of one-dimensional Lie subalgebras $Z$ such that $c_Z=c_Y$.

If it is also
\[
Y\,\mathcal{R}\,X\quad\hbox{or}\quad (Y,X)\in\mathcal{R},
\]
then we say that the two families $X$ and $Y$ are \emph{equivalent}.
\end{definition}

According to Definition~\ref{def:relationpfamily1D}, the relation $\mathcal{R}$  is clearly reflexive and transitive, but, in general, it is not necessarily symmetric, as will be shown in Example~\ref{exa:equiv1d}. In such a case, $\mathcal{R}$ is a so-called \emph{preorder}.

\begin{example}\label{exa:equiv1d}
Let $\mathcal{L}_4$ be a four-dimensional Lie algebra, 
\[
\mathcal{L}_4=\left\langle \Xi_1,\Xi_2,\Xi_3,\Xi_4\right\rangle, 
\]
with nonzero Lie brackets
\[
\left[\Xi_2,\Xi_4\right]=\Xi_1, \qquad \left[\Xi_3,\Xi_4\right]=\Xi_2.
\]
As the inner automorphism $\exp(t_1 \ad_{\Xi_1})$ is the identity morphism, let us write the matrices associated to inner automorphisms $A_k=\exp(t_k \ad_{\Xi_k})$, $k=2,3,4$:
\[
A_2 =
\begin{pmatrix}
1 & 0 & 0 & -t_2\\
0 & 1 & 0 & 0\\
0 & 0 & 1 & 0\\
0 & 0 & 0 & 1
\end{pmatrix},\qquad
A_3 =
\begin{pmatrix}
1 & 0 & 0 & 0\\
0 & 1 & 0 & -t_3\\
0 & 0 & 1 & 0\\
0 & 0 & 0 & 1
\end{pmatrix},\qquad
A_4 =
\begin{pmatrix}
1 & t_4 & \frac{t_4^2}{2} & 0\\
0 & 1 & t_4 & 0\\
1 & 0 & 1 & 0\\
0 & 0 & 0 & 1
\end{pmatrix}.
\]

Action of the automorphism $A_4$ on the $2$-family $\left(0,f^2 , f^3, 0\right)$ yields
\[
\left(f^2 t_4 + \frac{f^3t_4^2}{2},f^2+f^3 t_4, f^3, 0\right),
\]
which is a 3-family according to Definition~\ref{def:family1D}; thus, we can say
\[
\left(0,f^2,f^3,0\right)\ \mathcal{R}\ \left(f^1,f^2,f^3,0\right).
\]
Vice versa, it is
\[
\left(f^1 ,f^2 , f^3, 0\right) \stackrel{A_4}{\mapsto} \left(f^1 +f^2 t_4+\frac{f^3t_4^2}{2},f^2+f^3 t_4, f^3, 0\right), 
\]
and the image cannot be reduced to $\left(0,f^2,f^3,0\right)$ for all choices of $f^1$, $f^2$ and $f^3$, whence  
\[
\left(f^1,f^2,f^3,0\right)\;  \cancel{\mathcal{R}}\; \left(0,f^2,f^3,0\right).
\]
Moreover, choosing $\displaystyle t_4=-\frac{f^2}{f^3}$,  the 2-family $(0,f^2 , f^3, 0)$ is mapped by $A_4$
to $\displaystyle \left(-\frac{(f^2)^2}{2 f^3} ,0, f^3, 0\right)$, which is a $2$-family. Hence, we can write
\[
\left(0,f^2,f^3,0\right)\,\mathcal{R}\,\left(f^1, 0, f^3, 0\right). 
\]
Finally, as
\[
\left(f^1 ,0 , f^3, 0\right)\stackrel{A_4}{\mapsto} \left(f^1 +\frac{f^3t_4^2}{2},f^3 t_4, f^3, 0\right),
\]
we have
\[
\left(f^1,0,f^3,0\right)\,\mathcal{R}\,\left(f^1, f^2, f^3, 0\right), 
\]
but
\[
\left(f^1,0,f^3,0\right)\,\cancel{\mathcal{R}}\,\left(0, f^2, f^3, 0\right). 
\]
\end{example}

\begin{definition}
The families of one-dimensional Lie subalgebras are sorted according to the short-lexicographical ordering (slex). Let $\mathcal{L}$ be an $r$-dimensional Lie algebra and let 
\[
X =(f^1 s_1,f^2 s_2,\ldots,f^r s_r),\qquad
Y = (\tilde{f}^1 \tilde{s}_1, \tilde{f}^2 \tilde{s}_2,\ldots,\tilde{f}^r \tilde{s}_r) 
\]
be two families. It is
\[
X >_{\hbox{slex}} Y\quad \hbox{if} \quad
\left\{
\begin{aligned}
&\sum_{i=1}^r  s_i> \sum_{i=1}^r \tilde{s}_i, \\
&\hbox{or}\\
& \sum_{i=1}^r  s_i= \sum_{i=1}^r \tilde{s}_i \quad\hbox{and}\quad c_X > c_Y .
\end{aligned}
\right.
\]
\end{definition}

\begin{remark}
\label{rem:graph}
For an $r$-dimensional Lie algebra, all $2^r-1$ possible families of one-dimensional subalgebras, according to relation $\mathcal{R}$,  can be represented by means of a suitable \emph{directed} multigraph $\mathcal{G}(\mathcal{L})$ (this is because more than one automorphism can be such that the two families belong to the relation $\mathcal{R}$). The vertices correspond to the various families of one-dimensional Lie subalgebras, and the edges to the automorphisms connecting pairs of families belonging to 
$\mathcal{R}$. Indeed, we represent this multigraph as a graph by means of its adjacency matrix whose $(i,j)$th entry is 1 if the $i$th family is mapped by some automorphism to the $j$th family, and zero otherwise.
When we represent the graph corresponding to the set of $p$-families of one-dimensional Lie subalgebras, the label we append to each vertex $X$ is the position $\hbox{ind}_X$ in the list of ordered families.  
\end{remark}

\subsection{Higher dimensional Lie subalgebras}
Let us now define a $p$-family of $d$-dimensional ($1<d<r$) Lie subalgebras.

\begin{definition}
\label{def:familymultiD}
Let $\mathcal{L}$ be an $r$-dimensional Lie algebra spanned by $\{\Xi_1,\ldots,\Xi_r\}$, and let 
$f_k^\alpha$ ($k=1,\ldots,d$, $\alpha=1,\ldots,r$) be $d\cdot r$ real functions depending on some real variables belonging to the set 
$\mathcal{P}$.
A set of $d$ elements $({X}_1,\ldots,{X}_d)$, where
\[
{X}_k=\sum_{\alpha=1}^r f_k^\alpha s_{k,\alpha} \Xi_{\alpha},
\qquad k=1,\ldots,d,
\]
and $\mathbf{s}_k\equiv(s_{k,1},\ldots,s_{k,r}) \in \mathcal{S}_r$ for $k=1,\ldots,d$,
is called a $p$-family of $d$-dimensional Lie subalgebras of $\mathcal{L}$ if 
\begin{enumerate}
\item ${X}_k$ $(k=1,\ldots,d)$ is a $p_k$-family of one-dimensional Lie subalgebras, and $\displaystyle{p=\sum_{k=1}^d p_k}$; 
\item the matrix $\| f_k^\alpha t_{k,\alpha}\|$ has rank $d$; 
\item the rank of the Jacobian matrix of $\{f_k^{\alpha},\; k=1,\ldots,d,\; \alpha=1,\ldots,r\}$ 
with respect to the elements of $\mathcal{P}$ is equal to $p$;
\item the conditions
\[
\sum_{\alpha,\beta=1}^r f_i^\alpha s_{i,\alpha} f_j^\beta s_{j,\beta} C_{\alpha\beta}^\gamma = \sum_{k=1}^s \lambda_{ij}^k f_k^\gamma s_{k,\gamma},  \quad (i,j=1,\ldots, d, \gamma=1,\ldots,r)
\]
are satisfied for suitable constants $\lambda_{ij}^k$ whatever the values of the nonvanishing constants $f_k^\alpha$ are.
\end{enumerate}
\end{definition}

\begin{remark}
Definition~\ref{def:familymultiD} (in particular, item 4) implies a deep simplification of the conditions for the check that the set $\{X_1,\ldots,X_d\}$ is closed with respect to the Lie bracket; in fact, we need to determine the unknowns $\lambda_{ij}^k$ whatever the values of $f_k^\alpha$ are, and this requires only simple elementary linear algebra tools.

We are aware that Definition~\ref{def:familymultiD} in some cases may leave out some Lie subalgebras (see \cite{AmataOliveri2} for an example).
\end{remark}

\begin{remark}
Of course, because the basis of a $d$-dimensional Lie subalgebras of a Lie algebra $\mathcal{L}$ is not unique, 
we choose to adopt as the basis of a $p$-family of $d$-dimensional Lie subalgebras the one such that the matrix
\[
\left\|
\begin{array}{llll}
f_1^1 s_{1,1}  & f_1^2 s_{1,2} & \cdots  & f_1^r s_{1,r}\\
\vdots & \vdots & \vdots & \vdots\\ 
f_d^1 s_{d,1}  & f_d^2 s_{d,2} & \cdots  & f_d^r s_{d,r}
\end{array}
\right\|
\]
is in row reduced echelon form (RREF); once the basis for $\mathcal{L}$ has been assigned, this matrix represents the Lie subalgebra.
\end{remark}

\begin{definition}[Ordering of $p$--families of $d$--dimensional Lie subalgebras]
The $p$--families of $d$--dimensional Lie subalgebras are sorted according to the short-lexicogra\-phical ordering. Let $\mathcal{L}$ be an $r$--dimensional Lie algebra and let $X\equiv\{X_1,\ldots,X_d\}$ and 
$Y\equiv\{Y_1,\ldots,Y_d\}$ be a $p$--family and $q$--family of different $d$--dimensional Lie subalgebras, respectively. 
It is
\[
X >_{\hbox{slex}} Y \quad \hbox{if}\quad
\left\{
\begin{array}{l}
p > q,\\
\hbox{or}\\
p = q\quad\hbox{and}\quad X_k >_{\hbox{slex}}Y_k, \; \hbox{where}\; k =\hbox{min}\{i: c_{X_i}\neq c_{Y_i}\}. 
\end{array}
\right.
\]
\end{definition}

\begin{remark}
\label{remark:familyCoding}
Analogously to one-dimensional families of Lie subalgebras,
multidimensional $p$-families ${X}\equiv\{{X}_1,\ldots,{X}_d\} $ of dimension $d$, can be identified  by the tuple $c_{X}\equiv(c_{X_1},\ldots,c_{X_d})$.
\end{remark}

\begin{remark}
We want to clarify that, with Definition~\ref{def:family1D} and Definition~\ref{def:familymultiD}, we do not intend to deal with sets of subalgebras of $\mathcal{L}$ in the classical sense (that is, as the collection of subalgebras that vary with the variation of variables belonging to the set $\mathcal{P}$), but rather, a p-family is a self-consistent object. For example, if $\mathcal{L}$ is a Lie algebra of dimension $4$, we treat its $2$-family ${X} = (f^1, f^2, 0, 0)$ and its $2$-family ${Y} = ((f^1)^2, f^2, 0, 0)$ as the same object, and to us, they can be identified with the integer $c_X = c_Y = 3$. The same principle applies to multidimensional $p$-families. 
\end{remark}

For the sake of completeness, we illustrate the underlying ideas of the main algorithms yielding the automatic determination of optimal systems of an assigned $r$-dimensional Lie algebra, defined by its structure constants, or by giving its basis and the list of nonzero Lie brackets, or giving one realization in terms of vector fields or matrices (see \cite{AmataOliveri2} for more details). 

According to Definition~\ref{def:family1D}, we have $2^r -1$ possible one-dimensional $p$-families of Lie subalgebras, each identified by the tuple $X=(f^1 s_1,\ldots, f^r s_r)$. Hence, the action of an inner automorphism on a $p$-family can be transferred to the corresponding tuple. Furthermore, although such $p$-families can be identified  with the integers $c_X$, within the package they are sorted according to short-lexicographic (slex) ordering.  The position of a $p$-family in this ordering is used to label the vertex representing it in the oriented multigraph $\mathcal{G}(\mathcal{L})$. Then, the vertices of such a graph are the one-dimensional $p$-families of Lie subalgebras of $\mathcal{L}$, and the edges are given by the action of the inner automorphisms on the vertices.
 
More technically, let $X=\{f^1 s_1,\ldots, f^r s_r\}$ and $Y=\{g^1 s^\prime_1,\ldots, g^r s^\prime_r\}$ be two families of one-dimensional Lie subalgebras of $\mathcal{L}$. We already said that $X$ and $Y$ belong to the relation $\mathcal{R}$ if the action of some inner automorphism  $A$ on each element of $X$ produces a $p$-family $Z$ of Lie subalgebras such that $c_Z = c_Y$. During the computation, if it is found that $X\, \mathcal{R}\, Y$, the adjacency matrix of $\mathcal{G}(\mathcal{L})$ is updated and it is checked whether 
$Y\, \mathcal{R}\, X$ too. If $X\, \mathcal{R}\, Y$ only, in the graph the vertices corresponding to $X$ and $Y$ are connected by a directed edge, whereas if it is also $Y\, \mathcal{R}\, X$ by an
undirected edge. At the end of all these checks, the multigraph can be represented, and for each connected component a representative is selected: the vertex representing the smallest (in the slex ordering) $p$-family with maximal indegree. The set of these representatives yields the optimal system of families of one-dimensional Lie subalgebras. 

As far as higher dimensional $p$-families of Lie subalgebras are concerned, the reasoning is analogous, although some further observations are needed.

Let $V$ be a finite-dimensional vector space and let $W\subseteq V$ be a subspace.
We can consider basis vectors of $W$ and arrange them as rows in a matrix. Thus, 
$W$ is equal to the row space of the matrix built. Moreover, a matrix and its RREF have the same row space (they are row-equivalent), so $W$ is always spanned by rows of its associated matrix in RREF and such a matrix is unique.

Therefore, it is sufficient to consider multidimensional $p$-families of Lie subalgebras for which the corresponding matrices are in row reduced echelon form; these will be the \emph{candidates} that the algorithm will examine.

We observe that in the multidimensional case, the number of \textit{candidates} for $p$-families of Lie subalgebras depends not only on the dimension of the Lie algebra but also on its structure constants. 

As with the one-dimensional case, during the execution of the algorithm, starting from the smallest multidimensional $p$-family (with respect to the slex ordering), automorphisms are applied, and the links between $p$-families are saved in the adjacency matrix. In such a way it is possible to obtain the optimal system of families of subalgebras of dimension $d > 1$.

Finally, in both the one- and multi-dimensional case, the coefficients of the representatives of the optimal system can be rescaled if a scaling automorphism acting on them exists. In this case, we represent them with Greek letters which assume the values $\pm 1$, otherwise we represent them with Latin letters which assume arbitrary nonzero values (for further details see \cite{AmataOliveri2}).

\section{How to use \symbolie}
\label{sec:symbolie}

In this Section, we illustrate with a simple example how \symbolie can be used for recovering optimal systems of families of Lie subalgebras.  

To use \symbolie, open a Mathematica notebook and load the package by issuing 
\begin{verbatim}
<< "SymboLie.wl"
\end{verbatim}
So doing, we can specify the Lie algebra we want to analyze and start the computation.

Therefore, we clarify the notation used in \symbolie and compare the results with those obtained in \cite{PW}. 

\begin{example}
Let $\mathcal{L}_4$ be the Lie algebra $A_2\oplus 2 A_1$ (see \cite[Table II]{PW}) spanned by $\{\Xi_1,\Xi_2,\Xi_3,\Xi_4\}$ with only one nonzero commutator, say
\[
[\Xi_1,\Xi_2]=\Xi_2.
\]
This Lie algebra can be defined in \symbolie fixing the structure constants, say
\begin{verbatim}
sc = Table[0, {i,1,4}, {j,1,4}, {k,1,4}];
sc[[1,2,2]] = 1; sc[[2,1,-2]] = -1;
\end{verbatim}
or listing the basis of the Lie algebra and the nonzero commutators and computing the structure constants, say
\begin{verbatim}
basis = {e1,e2};
{{{e1,e2},{e2}}};
sc =  StructureConstants[basis,gens];
\end{verbatim}
Notice that whatever the symbols for the elements of the basis are, \symbolie will display the results using the symbols $\Xi_1$, $\Xi_2$, \ldots for the generators of the basis. 
The inner automorphisms are easily computed:
\begin{verbatim}
auto = InnerAutomorphisms[sc,t];
\end{verbatim}
The matrices $A_1$ and $A_2$ associated to the nontrivial inner automorphisms $\exp(t_1 \ad_{\Xi_1})$ and $\exp(t_2 \ad_{\Xi_2})$ read:
\[
A_1 =
\begin{pmatrix}
1 & 0 & 0 & 0\\
0 & \exp(t_1) & 0 & 0\\
0 & 0 & 1 & 0\\
0 & 0 & 0 & 1
\end{pmatrix},\
A_2 =
\begin{pmatrix}
1 & 0 & 0 & 0\\
-t_2 & 1 & 0 & 0\\
0 & 0 & 1 & 0\\
0 & 0 & 0 & 1
\end{pmatrix}.
\]
In \cite{PW}, the following one-dimensional optimal system (written using the notation of \symbolie) has been exhibited:
\begin{equation}\label{eq:4D_2_theta1}
\begin{aligned}
\Theta_{\mathcal{A}}^1 \equiv\  \{ & \{\Xi_2\}, \{\cos(\phi)\Xi_3+\sin(\phi)\Xi_4\}, \{\Xi_1 + x(\cos(\phi) \Xi_3 + \sin(\phi) \Xi_4)\},\\
& \{\Xi_2 + \epsilon(\cos(\phi)\Xi_3 + \sin(\phi)\Xi_4)\} \},
\end{aligned}
\end{equation}
with $0\leq \phi<\pi$, $x\in\mathbb{R}$ and $\epsilon= \pm 1$.
First, let us analyze $\{\cos(\phi)\Xi_3+\sin(\phi)\Xi_4\}$. If  $\sin(\phi)=0$ then we obtain the subalgebra $\{\Xi_3\}$; if $\cos(\phi)=0$ then we obtain $\{\Xi_4\}$; otherwise, we can write $\{\Xi_3+\tan(\phi)\Xi_4\}$ with $\tan(\phi)\in \mathbb{R}^\star=\mathbb{R}\setminus\{0\}$. 
Similarly, with regard to $\{\Xi_1 + x(\cos(\phi) \Xi_3 + \sin(\phi) \Xi_4)\}$, $x\cos(\phi)$ and $x\sin(\phi)$ assume every real values. Hence, if $x= 0$ we get $\{\Xi_1\}$, else $\{\Xi_1+x\Xi_3\}$, $\{\Xi_1+x\Xi_4\}$ and $\{\Xi_1+x\Xi_3+y\Xi_4\}$ with $x,y\in\mathbb{R}^\star$. Finally, from $\{\Xi_2 + \epsilon(\cos(\phi)\Xi_3 + \sin(\phi)\Xi_4)\}$ we obtain $\{\Xi_2 + \epsilon\Xi_3\}$, $\{\Xi_2 + \epsilon\Xi_4\}$ and the last $\{\Xi_2 + \epsilon(\cos(\phi)\Xi_3 + \sin(\phi)\Xi_4)\}$ when $\phi\in ]0,\pi/2[\cup ]\pi/2, \pi[$.

Then, the one-dimensional optimal system exhibited in \cite{PW},  written in extended form, is
\begin{align*}
\Theta_{\mathcal{A}}^1 \equiv\ \{&\{\Xi_1\}, \{\Xi_2\},  \{\Xi_3\},\{\Xi_4\}, \{\Xi_3+x\Xi_4\},\{\Xi_1 + x \Xi_3\}, \{\Xi_1 + x \Xi_4\},\\
& \{\Xi_1 + x \Xi_3 + y \Xi_4\}, \{\Xi_2+\epsilon \Xi_3\},\{\Xi_2+\epsilon\Xi_4\},
 \{\Xi_2 + \epsilon(\cos(\phi)\Xi_3 + \sin(\phi)\Xi_4)\} \},
\end{align*}
with $\phi\in ]0,\pi[\setminus\{\pi/2\}$, $x,y\in\mathbb{R}^\star$ and $\epsilon= \pm 1$.

The program \symbolie returns 
\begin{align*}
\Psi_{\mathcal{A}}^1 \equiv\ & \{\{\Xi_1\}, \{\Xi_2\},  \{\Xi_3\},\{\Xi_4\}, \{\Xi_3+a_1\Xi_4\},\{\Xi_1 + a_1 \Xi_3\}, \{\Xi_1 + a_1\Xi_4\},\\
&\{\Xi_1 + a_1 \Xi_3 + a_2 \Xi_4\},\{\Xi_2+\alpha_1 \Xi_3\},\{\Xi_2+\alpha_1\Xi_4\}, \{\Xi_2 + \alpha_1\Xi_3 + a_1\Xi_4\} \},
\end{align*}
with $a_1,a_2\in \mathbb{R}^\star$ and $\alpha_1=\pm 1$.

The cardinality of the one-dimensional optimal system reported in \cite{PW} is the same as that of the optimal system computed by \symbolie. The two sets are equal except for one element. Anyway, they are equivalent. Indeed, let us consider
\[
K_1 = \{\Xi_2 + \alpha_1(\cos(\phi)\Xi_3 + \sin(\phi)\Xi_4)\},\ \text{with}\ \alpha_1 = \pm 1\ \text{and}\ \phi\in ]0,\pi[\setminus\{\pi/2\}.
\]
Applying the inner automorphism $A_1$, it is
\[
(0, 1, \alpha_1\cos(\phi), \alpha_1\sin(\phi))^T \stackrel{A_1}{\mapsto}  (0, \exp(-t_1), \alpha_1\cos(\phi), \alpha_1\sin(\phi)\Xi_4).
\]
Let $K_2$ be the subalgebra generated by it, with $t_1 = \log(1/\cos(\phi))$. Thus, one has
\[
K_2 = \{\Xi_2 + \alpha_1\Xi_3+\alpha_1\tan(\phi)\Xi_4\},
\]
and assuming $a_1=\alpha_1\tan(\phi)\in\mathbb{R}^\star$, we prove the equivalence.

A similar argument applies in the case of the two-dimensional optimal system listed in \cite{PW}, which in extended form can be written as
\begin{align*}
\Theta_{\mathcal{A}}^2 \equiv\ \{&\{\Xi_1,\Xi_2\}, \{\Xi_1 + x \Xi_3, \Xi_2\}, \{\Xi_1+x\Xi_4, \Xi_2\}, \{\Xi_1 + x\cos(\phi)\Xi_3 + x\sin(\phi)\Xi_4, \Xi_2\},\\
& \{\Xi_1, \Xi_3\}, \{\Xi_1,\Xi_4\}, \{\Xi_1, \sin(\phi)\Xi_3 - \cos(\phi)\Xi_4\}, \{\Xi_1 + x\Xi_3, \Xi_4\}, \{\Xi_1 + x\Xi_4, \Xi_3\},\\
& \{\Xi_2+\epsilon\Xi_3, \Xi_4\}, \{\Xi_2+\epsilon\Xi_4, \Xi_3\}, \{\Xi_3,\Xi_4\},\{\Xi_2, \Xi_3\}, \{\Xi_2,\Xi_4\},\\
& \{\Xi_2, \sin(\phi)\Xi_3 - \cos(\phi)\Xi_4\}, \{\Xi_1 + x\cos(\phi)\Xi_3+x\sin(\phi)\Xi_4, \sin(\phi)\Xi_3-\cos(\phi)\Xi_4\},\\
& \{\Xi_2 + \epsilon(\cos(\phi)\Xi_3+\sin(\phi)\Xi_4), \sin(\phi)\Xi_3-\cos(\phi)\Xi_4\}
 \},
\end{align*}
with $\phi\in ]0,\pi[\setminus\{\pi/2\}$, $x\in\mathbb{R}^\star$ and $\epsilon= \pm 1$.

Using \symbolie, we obtain
\begin{align*}
\Psi^2_{\mathcal{A}} \equiv\ \{&\{\Xi_1,\Xi_2\}, \{\Xi_1 + a_1 \Xi_3, \Xi_2\}, \{\Xi_1+a_1\Xi_4,\Xi_2\}, \{\Xi_1 + a_1\Xi_3 + a_2\Xi_4, \Xi_2\},\\
&  \{\Xi_1, \Xi_3\}, \{\Xi_1,\Xi_4\}, \{\Xi_1, \Xi_3 + a_1\Xi_4\}, \{\Xi_1 + a_1\Xi_3, \Xi_4\}, \{\Xi_1 + a_1\Xi_4, \Xi_3\},\\
& \{\Xi_2+\alpha_1\Xi_3, \Xi_4\}, \{\Xi_2+\alpha_1\Xi_4, \Xi_3\}, \{\Xi_3,\Xi_4\},\{\Xi_2, \Xi_3\}, \{\Xi_2,\Xi_4\},\\
& \{\Xi_2, \Xi_3 +a_1\Xi_4\},
\{ \Xi_1+a_1\Xi_4, \Xi_3 + a_2\Xi_4\}, \{\Xi_2+\alpha_1\Xi_4, \Xi_3 + a_1\Xi_4\}
 \},
\end{align*}
with $a_1,a_2\in \mathbb{R}^\star$ and $\alpha_1=\pm 1$.

Also in this case the cardinalities of the two optimal systems coincide, and the representatives of $\Theta_{\mathcal{A}}^2$ and $\Psi^2_{\mathcal{A}}$ are all the same except the last two.

Concerning $\{\Xi_1 + x\cos(\phi)\Xi_3+x\sin(\phi)\Xi_4, \sin(\phi)\Xi_3-\cos(\phi)\Xi_4\}$, we observe that after a row reduction it can be written as
\[
\{\Xi_1 + x\csc(\phi)\Xi_4, \Xi_3 - \cot(\phi)\Xi_4\},
\] 
and since $x$ and $\phi$ are arbitrary, we can set $a_1 = x\csc(\phi)$ and $a_2 = -\cot(\phi)$, obtaining
\[
\{\Xi_1 + a_1\Xi_4, \Xi_3 + a_2\Xi_4\}.
\]
Similarly, for $\{\Xi_2 + \epsilon(\cos(\phi)\Xi_3+\sin(\phi)\Xi_4), \sin(\phi)\Xi_3-\cos(\phi)\Xi_4\}$ after a row reduction we obtain
\[
\{\Xi_2 + \epsilon\csc(\phi)\Xi_4, \Xi_3 -\cot(\phi)\Xi_4\}.
\]
Moreover, applying $A_1$ with $t_1 = \log(\sin(\phi))$, we get
\[
\{\Xi_2 + \epsilon\Xi_4, \Xi_3 - \cot(\phi)\Xi_4\},
\]
and assuming $\alpha_1=\epsilon$ and $a_1 = -\cot(\phi)$ we obtain $\{\Xi_2 + \alpha_1\Xi_4, \Xi_3+a_1\Xi_4\}$.

Finally, the three-dimensional optimal systems listed in \cite{PW} and computed by \symbolie are 
\begin{align*}
\Theta_{\mathcal{A}}^3 \equiv\ \{& \{\Xi_1,\Xi_3,\Xi_4\}, \{\Xi_2, \Xi_3, \Xi_4\}, \{\Xi_1, \Xi_3, \Xi_2\}, \{\Xi_1, \Xi_4, \Xi_2\},\\
& \{\Xi_1, \sin(\phi)\Xi_3-\cos(\phi)\Xi_4,\Xi_2\}, \{\Xi_1+x\Xi_3, \Xi_4, \Xi_2\},
 \{\Xi_1+x\Xi_4, \Xi_3, \Xi_2\},\\
 & \{\Xi_1+x\cos(\phi)\Xi_3+x\sin(\phi)\Xi_4, \sin(\phi)\Xi_3-\cos(\phi)\Xi_4,\Xi_2\}
 \},
\end{align*}
with $\phi\in ]0,\pi[\setminus\{\pi/2\}$ and $x\in\mathbb{R}^\star$, and
\begin{align*}
\Psi_{\mathcal{A}}^3 \equiv\ \{& \{\Xi_1,\Xi_3,\Xi_4\}, \{\Xi_2, \Xi_3, \Xi_4\}, \{\Xi_1, \Xi_2, \Xi_3\}, \{\Xi_1, \Xi_2, \Xi_4\},\\
& \{\Xi_1, \Xi_2,\Xi_3+a_1\Xi_4\}, \{\Xi_1+a_1\Xi_3, \Xi_2, \Xi_4\},
 \{\Xi_1+a_1\Xi_4, \Xi_2, \Xi_3\},\\
 & \{\Xi_1+a_1\Xi_4, \Xi_2, \Xi_3+a_2\Xi_4\}
 \},
\end{align*}
with $a_1,a_2\in \mathbb{R}^\star$, respectively.

The representatives of the both three-dimensional optimal systems coincide. In particular, after the row reduction the last representative of $\Theta_{\mathcal{A}}^3$ becomes 
\[
\{\Xi_2,\Xi_1+x\csc(\phi)\Xi_4, \Xi_3-\cot(\phi)\Xi_4\},
\]
and, assuming $a_1 = x \csc(\phi)$ and $a_2=-\cot(\phi)$, we obtain $\{\Xi_2,\Xi_1+a_1\Xi_4, \Xi_3+a_2\Xi_4\}$; thus, the two three-dimensional optimal systems coincide.
\end{example}

\section{Analysis of three-dimensional real Lie algebras}
\label{sec:3D}
The optimal systems of Lie subalgebras for real three-dimensional Lie algebras were listed in~\cite[Table~I]{PW}. All such optimal systems coincide with those computed by \symbolie, except the algebra $A_{3,8}$. Here, we analyze this case.

\begin{algebra}[$\mathbf{A_{3,8}}$]
Let $\mathcal{L}_3$ be the Lie algebra spanned by $\{\Xi_1,\Xi_2,\Xi_3\}$ with the nonzero commutators:
\[
[\Xi_1,\Xi_2]=\Xi_1,\quad [\Xi_2,\Xi_3]=\Xi_3,\quad [\Xi_3,\Xi_1]=2\Xi_2.
\]
The matrices associated to the inner automorphisms $\exp(t_k \ad_{\Xi_k})$, $k=1,2,3$, are
\[
A_1 =
\begin{pmatrix}
1 & -t_1 & -t_1^2\\
0 & 1 & 2t_1\\
0 & 0 & 1
\end{pmatrix},\
A_2 =
\begin{pmatrix}
\exp(t_2) & 0 & 0\\
0 & 1 & 0\\
0 & 0 & \exp(-t_2)
\end{pmatrix},\
A_3 =
\begin{pmatrix}
1 & 0 & 0\\
-2t_3 & 1 & 0\\
-t_3^2 & t_3 & 1
\end{pmatrix}.
\]
Let us denote by $\mathcal{A}$ the group generated by $\{A_1,A_2,A_3\}$.

In \cite{PW}, the optimal system
\begin{align*}
\Theta_{\mathcal{A}}^1\equiv \{\{\Xi_1\}, \{\Xi_2\},  \{\Xi_1+\Xi_3\}\},\quad \Theta_{\mathcal{A}}^2\equiv \{\{\Xi_1,\Xi_2\}\}
\end{align*}
is listed; on the contrary,  \symbolie produces
 \[
\Psi_{\mathcal{A}}^1\equiv \{\{\Xi_1\}, \{\Xi_2\}, \{\Xi_1+\alpha_1 \Xi_3\}\},\quad \Psi_{\mathcal{A}}^2\equiv \{\{\Xi_1,\Xi_2\}\}.
\]
As far as the two-dimensional optimal system is concerned, we have $\Theta^2_{\mathcal{A}}=\Psi^2_{\mathcal{A}}$.

On the other hand, the one-dimensional optimal systems $\Theta^1_{\mathcal{A}}$ and $\Psi^1_{\mathcal{A}}$ differ for the third representative. As remarked in Section~\ref{sec:preliminaries}, the subalgebra ${\Xi_1+\Xi_3}$ is not considered in \symbolie (it does not form a $2$-family); therefore, we must refer to the corresponding $2$-family $X={\Xi_1 + a_1\Xi_3}$. Furthermore, since the coefficient $a_1$ in the $2$-family $X$ can be rescaled, it is represented by the Greek letter $\alpha$, that can assume the values $\pm 1$. The one-dimensional optimal system computed by \symbolie is represented graphically in Figure~\ref{fig:graph3D10}.

In Table~\ref{tab:optimal3d}, we list the optimal systems of all nonisomorphic three-dimensional real Lie algebras; the results are produced in a few minutes by \symbolie (the file with the computation can be found at the url \url{https://mat521.unime.it/oliveri}).

\begin{figure}[h]
\label{fig:graph3D10}
\centering
\includegraphics[width=.4\textwidth]{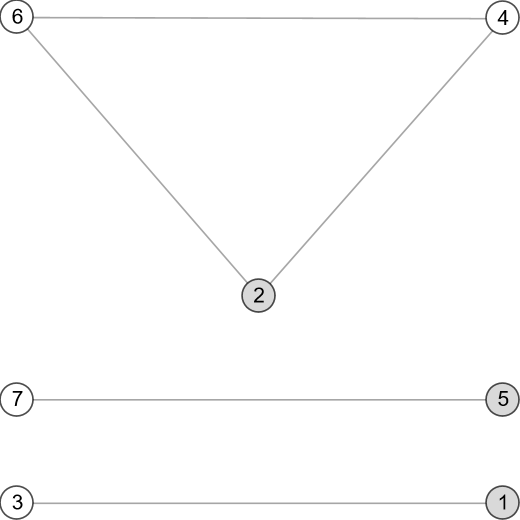}
\caption{One-dimensional optimal system for the Lie algebra $A_{3,8}$ computed by \symbolie. \footnotesize $1\to \{\Xi_1\}$, $2\to \{\Xi_2\}$, $3\to \{\Xi_3\}$, $4\to \{\Xi_1+a_1 \Xi_2\}$, $5\to \{\Xi_1+a_1 \Xi_3\}$, $6\to \{\Xi_2+a_1 \Xi_3\}$, $7\to \{\Xi_1+a_1 \Xi_2+a_2 \Xi_3\}$.}
\end{figure}
\end{algebra}
\begin{table}
\caption{\label{tab:optimal3d}Optimal systems of real Lie algebras of dimension 3.
The constants $a_1,a_2,a_3,a_4\in \mathbb{R}$ are nonvanishing, and $\alpha_1=\pm 1$.}
\label{tab:optimal3D}
{\smallsize
\begin{tabular}{p{1.5cm}p{4cm}p{3.8cm}p{4cm}}
\noalign{\smallskip}\hline\noalign{\smallskip}
\textbf{Algebra} & \textbf{Nonzero Lie brackets} & \textbf{1D Optimal System} & \textbf{2D Optimal System}\\
\noalign{\smallskip}\hline\noalign{\smallskip}
$3A_1$ & &
\begin{minipage}{3cm}
$\{\Xi_1\}$,  $\{\Xi_2\}$, $\{\Xi_3\}$,\\
$\{\Xi_1 + a_1\Xi_2\}$,\\
$\{\Xi_1 + a_1\Xi_3\}$,\\
$\{\Xi_2 + \alpha_1\Xi_3\}$ \\
$\{\Xi_1+a_1\Xi_2 + a_2\Xi_3\}$ 
\end{minipage} & 
\begin{minipage}{4cm}
$\{\Xi_1, \Xi_2\}$, $\{\Xi_1, \Xi_3\}$, $\{\Xi_2, \Xi_3\}$ \\
$\{\Xi_1, \Xi_2+a_1\Xi_3\}$\\
$\{\Xi_1+a_1 \Xi_2,\Xi_3\}$\\
$\{\Xi_1+a_1 \Xi_3,\Xi_2\}$\\
$\{\Xi_1+a_1 \Xi_3,\Xi_2+a_2\Xi_3\}$\\
\end{minipage}\\
\noalign{\smallskip}\hline\noalign{\smallskip}
$A_1\oplus A_2$ &
\begin{minipage}{3cm}
$[\Xi_1,\Xi_2]=\Xi_2$  
\end{minipage} & 
\begin{minipage}{3cm}
$\{\Xi_1\}$,  $\{\Xi_2\}$, $\{\Xi_3\}$,\\
$\{\Xi_1 + a_1\Xi_3\}$,\\
$\{\Xi_2 + \alpha_1\Xi_3\}$ 
\end{minipage} & 
\begin{minipage}{4cm}
$\{\Xi_1, \Xi_2\}$, $\{\Xi_1, \Xi_3\}$, $\{\Xi_2, \Xi_3\}$ \\
$\{\Xi_1+a_1\Xi_3, \Xi_2\}$
\end{minipage}\\
\noalign{\smallskip}\hline\noalign{\smallskip}
$A_{3,1}$ &
\begin{minipage}{3cm}
$[\Xi_2,\Xi_3]=\Xi_1$  
\end{minipage} & 
\begin{minipage}{3cm}
$\{\Xi_1\}$,  $\{\Xi_2\}$, $\{\Xi_3\}$,\\
$\{\Xi_2 + a_1\Xi_3\}$ 
\end{minipage} & 
\begin{minipage}{4cm}
$\{\Xi_1, \Xi_2\}$, $\{\Xi_1, \Xi_3\}$, \\
$\{\Xi_1, \Xi_2 + a_1\Xi_3\}$
\end{minipage}\\
\noalign{\smallskip}\hline\noalign{\smallskip}
$A_{3,2}$ &
\begin{minipage}{3cm}
$[\Xi_1,\Xi_3]=\Xi_1$, \\
$[\Xi_2,\Xi_3]=\Xi_1+\Xi_2$ 
\end{minipage} &
\begin{minipage}{3cm}
$\{\Xi_1\}$, $\{\Xi_2\}$, $\{\Xi_3\}$  
\end{minipage} &
\begin{minipage}{4cm}
$\{\Xi_1, \Xi_2\}, \{\Xi_1, \Xi_3\}$
\end{minipage} \\
\noalign{\smallskip}\hline\noalign{\smallskip}
$A_{3,3}$ &
\begin{minipage}{3cm}
$[\Xi_1,\Xi_3]=\Xi_1$, \\
$[\Xi_2,\Xi_3]=\Xi_2$ 
\end{minipage} & 
\begin{minipage}{3cm}
$\{\Xi_1\}$, $\{\Xi_2\}$, $\{\Xi_3\}$, \\
$\{\Xi_1 + a_1\Xi_2\}$ 
\end{minipage} & 
\begin{minipage}{4cm}
$\{\Xi_1, \Xi_2\}$, $\{\Xi_1, \Xi_3\}$, $\{\Xi_2, \Xi_3\}$, \\
$\{\Xi_1 + a_1\Xi_2, \Xi_3\}$
\end{minipage}\\
\noalign{\smallskip}\hline\noalign{\smallskip}
$A_{3,4}$ &
\begin{minipage}{3cm}
$[\Xi_1,\Xi_3]=\Xi_1$,\\ 
$[\Xi_2,\Xi_3]=-\Xi_2$ 
\end{minipage} & 
\begin{minipage}{3cm}
$\{\Xi_1\}$, $\{\Xi_2\}$, $\{\Xi_3\}$, \\
$\{\Xi_1 + \alpha_1\Xi_2\}$ 
\end{minipage} & 
\begin{minipage}{4cm}
$\{\Xi_1, \Xi_2\}$, $\{\Xi_1, \Xi_3\}$, $\{\Xi_2, \Xi_3\}$ 
\end{minipage}\\
\noalign{\smallskip}\hline\noalign{\smallskip}
$A_{3,5}^a$ &
\begin{minipage}{3cm}
$[\Xi_1,\Xi_3]=\Xi_1$,\\
$[\Xi_2,\Xi_3]=a \Xi_2$ \\
$(0<|a|<1)$ 
\end{minipage} &
\begin{minipage}{3cm}
$\{\Xi_1\}$, $\{\Xi_2\}$, $\{\Xi_3\}$, \\
$\{\Xi_1 + \alpha_1\Xi_2\}$ 
\end{minipage} & 
\begin{minipage}{4cm}
$\{\Xi_1, \Xi_2\}$, $\{\Xi_1, \Xi_3\}$, 
$\{\Xi_2, \Xi_3\}$
\end{minipage}\\ 
\noalign{\smallskip}\hline\noalign{\smallskip}
$A_{3,6}$ &
\begin{minipage}{3cm}
$[\Xi_1,\Xi_3]=-\Xi_2$, \\
$[\Xi_2,\Xi_3]=\Xi_1$ 
\end{minipage} & 
\begin{minipage}{3cm}
$\{\Xi_1\}$, $\{\Xi_3\}$  
\end{minipage} &
\begin{minipage}{4cm}
$\{\Xi_1, \Xi_2\}$ 
\end{minipage} \\ 
\noalign{\smallskip}\hline\noalign{\smallskip}
$A_{3,7}^a$ &
\begin{minipage}{3cm}
$[\Xi_1,\Xi_3]=a\Xi_1-\Xi_2$, \\
$[\Xi_2,\Xi_3]=\Xi_1+a\Xi_2$\\
$(a>0)$
\end{minipage} & 
\begin{minipage}{3cm}
$\{\Xi_1\}$, $\{\Xi_3\}$ 
\end{minipage} &
\begin{minipage}{4cm}
$\{\Xi_1, \Xi_2\}$ 
\end{minipage} \\
\noalign{\smallskip}\hline\noalign{\smallskip}
$A_{3,8}$ &
\begin{minipage}{3cm}
$[\Xi_1,\Xi_2]=\Xi_1$, \\
$[\Xi_1,\Xi_3]=-2\Xi_2,$\\
$[\Xi_2,\Xi_3]=\Xi_3$, \\
\end{minipage} & 
\begin{minipage}{3cm}
$\{\Xi_1\}$, $\{\Xi_2\}$, \\
$\{\Xi_1 + \alpha_1\Xi_3\}$ 
\end{minipage} &
\begin{minipage}{4cm}
$\{\Xi_1, \Xi_2\}$ 
\end{minipage} \\ 
\noalign{\smallskip}\hline\noalign{\smallskip}
$A_{3,9}$ &
\begin{minipage}{3cm}
$[\Xi_1,\Xi_2]=\Xi_3$,\\ 
$[\Xi_1,\Xi_3]=-\Xi_2$, \\
$[\Xi_2,\Xi_3]=\Xi_1$ 
\end{minipage} & 
\begin{minipage}{3cm}
$\{\Xi_1\}$ 
\end{minipage} \\
\noalign{\smallskip}\hline
\end{tabular}
}
\end{table}

\section{Analysis of four-dimensional Lie algebras}
\label{sec:4D}
Optimal systems of subalgebras for real four-dimensional Lie algebras have been determined in~\cite{PW}.
Similarly to the three-dimensional case, we analyze those algebras where there are some differences between the optimal systems computed by \symbolie and the ones listed in \cite[Table II]{PW}.

\begin{algebra}[$\mathbf{2A_2}$]
Let $\mathcal{L}_4$ be the Lie algebra spanned by $\{\Xi_1,\Xi_2,\Xi_3,\Xi_4\}$ with the nonzero commutators
\[
[\Xi_1,\Xi_2]=\Xi_2,\quad [\Xi_3,\Xi_4]=\Xi_4.
\]

The one-dimensional optimal system computed by \symbolie coincides with the one given in \cite{PW}. It is represented graphically in Figure~\ref{fig:graph4D3}.

\begin{figure}[h]
\centering
\includegraphics[width=.4\textwidth]{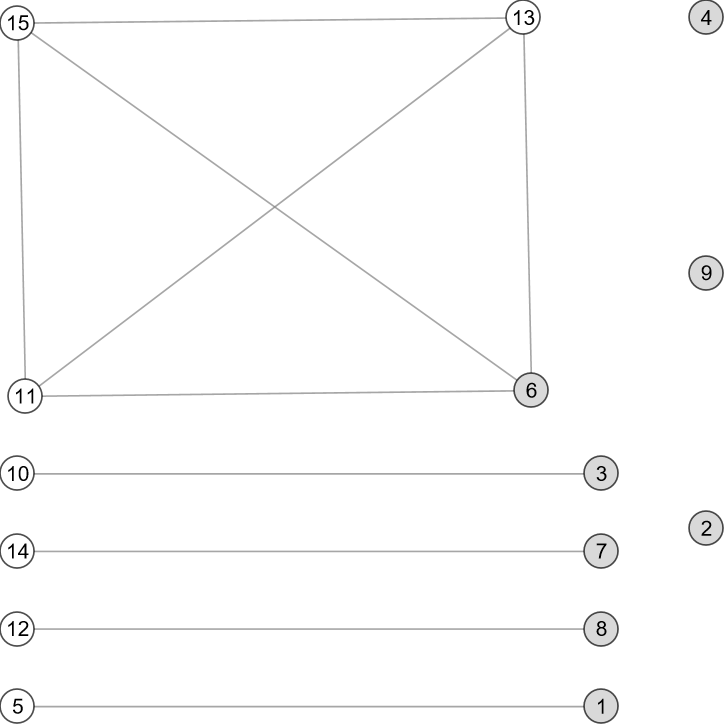}
\caption{One-dimensional optimal system for the Lie algebra $2A_2$ computed by \symbolie. \footnotesize\label{fig:graph4D3} $1\to \{\Xi_1\}$, $2\to \{\Xi_2\}$,$3\to \{\Xi_3\}$,$4\to \{\Xi_4\}$,$5\to \{\Xi_1+a_1\Xi_2\}$,$6\to \{\Xi_1+a_1 \Xi_3\}$,$7\to \{\Xi_2+a_1 \Xi_3\}$,$8\to \{\Xi_1+a_1 \Xi_4\}$,$9\to \{\Xi_2+a_1 \Xi_4\}$,$10\to \{\Xi_3+a_1 \Xi_4\}$,$11\to \{\Xi_1+a_1 \Xi_2+a_2 \Xi_3\}$,$12\to \{\Xi_1+a_1 \Xi_2+a_2 \Xi_4\}$,$13\to \{\Xi_1+a_1 \Xi_3+a_2 \Xi_4\}$,$14\to \{\Xi_2+a_1 \Xi_3+a_2 \Xi_4\}$,$15\to \{\Xi_1+a_1 \Xi_2+a_2 \Xi_3+a_3 \Xi_4\}$}
\end{figure}

Instead, concerning the two-dimensional  optimal systems, we have
\[
\Theta_{\mathcal{A}}^2 = \Psi^2_A \cup \{\{\Xi_1+\Xi_3, \Xi_2+\epsilon\Xi_4\}\},\quad \text{with}\ \epsilon = \pm 1.
\]

We observe that the representative in the set difference of the two systems is not a $p$-family. Moreover,  $\{\Xi_1+a_1\Xi_3, \Xi_2+a_2\Xi_4\}$, from which the missing subalgebra could arise, is not a $p$-family, because it does not satisfy, in particular, the condition $4$ of Definition~\ref{def:familymultiD} (for more details see \cite{AmataOliveri2}). Therefore, it cannot be analyzed during the calculation of the optimal system by \texttt{SymboLie}.

The three-dimensional optimal system computed by \symbolie is  
\begin{align*}
\Theta_{\mathcal{A}}^3 \equiv\ \{& \{\Xi_1,\Xi_2, \Xi_3\}, \{\Xi_1,\Xi_2, \Xi_4\},  \{\Xi_1, \Xi_3, \Xi_4\}, \{\Xi_2, \Xi_3,\Xi_4\}, \{\Xi_1+a_1\Xi_3, \Xi_2,\Xi_4\}\},
\end{align*}
with $a_1\in\mathbb{R}^\star$. The dissimilarity with respect to the optimal system in \cite{PW} consists in the $p$-family $\{\Xi_1+a_1\Xi_3, \Xi_2,\Xi_4\}$. In \cite{PW}, there are the special subalgebras $A_{3,3}$, $A_{3,4}$ and $A_{3,5}^a$ (under particular conditions) that represent the aforementioned $p$-family.
\end{algebra}

\begin{algebra}[$\mathbf{A_{3,6}\oplus A_1}$]
Let $\mathcal{L}_4$ be the Lie algebra spanned by $\{\Xi_1,\Xi_2,\Xi_3,\Xi_4\}$ with the nonzero commutators
\[
[\Xi_1,\Xi_3]=-\Xi_2,\quad [\Xi_2,\Xi_3]=\Xi_1.
\]
Let us write the matrices associated to inner automorphisms $\exp(t_k \ad_{\Xi_k})$, $k=1,2,3$, say
\[
A_1 =
\begin{pmatrix}
1 & 0 & 0 & 0\\
0 & 1 & t_1 & 0\\
0 & 0 & 1 & 0\\
0 & 0 & 0 & 1
\end{pmatrix},\
A_2 =
\begin{pmatrix}
1 & 0 & -t_2 & 0\\
0 & 1 & 0 & 0\\
0 & 0 & 1 & 0\\
0 & 0 & 0 & 1
\end{pmatrix},\
A_3 =
\begin{pmatrix}
\cos t_3 & \sin t_3 & 0 & 0\\
-\sin t_3 & \cos t_3 & 0 & 0\\
0 & 0 & 1 & 0\\
0 & 0 & 0 & 1
\end{pmatrix}.
\]

The optimal systems of order 1 and 3 in \cite{PW} coincide with those computed by \symbolie.
\begin{figure}[h]
\centering
\includegraphics[width=.4\textwidth]{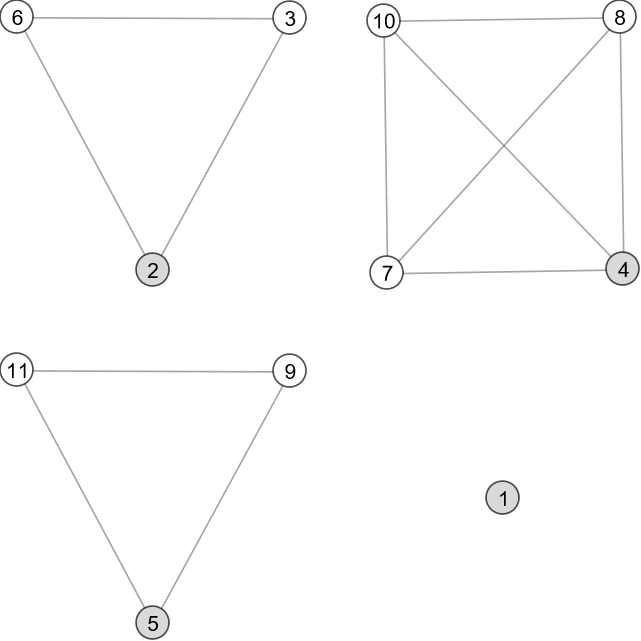}
\caption{Two-dimensional optimal system for the Lie algebra $A_{3,6}\oplus A_1$ computed by \texttt{SymboLie}. \footnotesize\label{fig:graph4D9} $1\to \{\Xi_1,\Xi_2\}$, $2\to \{\Xi_1,\Xi_4\}$, $3\to \{\Xi_2,\Xi_4\}$, $4\to \{\Xi_3,\Xi_4\}$, $5\to \{\Xi_1, \Xi_2+a_1 \Xi_4\}$, $6\to \{\Xi_1+a_1 \Xi_2,\Xi_4\}$, $7\to \{\Xi_1+a_1 \Xi_3,\Xi_4\}$, $8\to \{\Xi_2+a_1 \Xi_3,\Xi_4\}$, $9\to \{\Xi_1+a_1 \Xi_4,\Xi_2\}$, $10\to \{\Xi_1+a_1 \Xi_2+a_2 \Xi_3,\Xi_4\}$, $11\to \{\Xi_1+a_1 \Xi_4,a_2 \Xi_2+a_3 \Xi_4\}$}
\end{figure}
From the graph in Figure~\ref{fig:graph4D9}, it can be seen that two-dimensional optimal systems have different representatives of the same connected component. Indeed, the two-dimensional optimal systems in \cite{PW} and \symbolie are
\begin{align*}
\Theta_{\mathcal{A}}^2 \equiv\ \{& \{\Xi_1,\Xi_2\}, \{\Xi_1, \Xi_4\},  \{\Xi_1+x\Xi_4, \Xi_2\}, \{\Xi_3, \Xi_4\}\},\quad \text{and}\\
\Psi_\mathcal{{A}}^2 \equiv\ \{& \{\Xi_1,\Xi_2\}, \{\Xi_1, \Xi_4\},  \{\Xi_1, \Xi_2+a_1\Xi_4\}, \{\Xi_3, \Xi_4\}\},
\end{align*}
respectively.

Let us show that the family of subalgebras $\{\Xi_1,\Xi_2+a_1\Xi_4\}$ is equivalent to the family of subalgebras $\{\Xi_1+x\Xi_4, \Xi_2\}$ via the inner automorphism $A_3$. In fact, it is
\begin{align*}
(1,0,0,0) \stackrel{A_3}{\mapsto} (\cos (t_3), -\sin (t_3), 0, 0) \quad \text{and}\quad 
 (0,1,0,a_1)\stackrel{A_3}{\mapsto}  (\sin (t_3), \cos (t_3), 0, a_1).
\end{align*}
Choosing $t_3=\pi/2$, we immediately obtain the representative of $\Theta_{\mathcal{A}}^2$, and thus the optimal systems coincide.
\end{algebra}

\begin{algebra}[$\mathbf{A_{3,7}^a \oplus A_1}$]
Let $\mathcal{L}_4$ be the Lie algebra spanned by $\{\Xi_1,\Xi_2,\Xi_3,\Xi_4\}$ with the nonzero commutators
\[
[\Xi_1,\Xi_3]=a\Xi_1-\Xi_2,\quad [\Xi_2,\Xi_3]=\Xi_1+a\Xi_2,
\]
with $a > 0$.

The matrices associated to the inner automorphisms $\exp(t_k \ad_{\Xi_k})$, $k=1,2,3$, are
\begin{gather*}
A_1 =
\begin{pmatrix}
1 & 0 & -at_1 & 0\\
0 & 1 & t_1 & 0\\
0 & 0 & 1 & 0\\
0 & 0 & 0 & 1
\end{pmatrix},\
A_2 =
\begin{pmatrix}
1 & 0 & -t_2 & 0\\
0 & 1 & -t_2 & 0\\
0 & 0 & 1 & 0\\
0 & 0 & 0 & 1
\end{pmatrix},\\
A_3 =
\begin{pmatrix}
\exp(at_3)\cos t_3 & \exp(at_3)\sin t_3 & 0 & 0\\
-\exp(at_3)\sin t_3 & \exp(at_3)\cos t_3 & 0 & 0\\
0 & 0 & 1 & 0\\
0 & 0 & 0 & 1
\end{pmatrix}.
\end{gather*}

The optimal systems of dimension 1 and 3 coincide. As in the previous case, looking at the two-dimensional optimal system, 
\symbolie returns a representative distinct from the one obtained by Patera and Winternitz. Indeed, the two-dimensional optimal system found in \cite{PW} is
\begin{equation*}
\begin{aligned}
\Theta_{\mathcal{A}}^2 \equiv\ \{& \{\Xi_1,\Xi_2\}, \{\Xi_1, \Xi_4\},  \{\Xi_1+x\Xi_4, \Xi_2\}, \{\Xi_3, \Xi_4\}\},
\end{aligned}
\end{equation*}
and the one computed by \symbolie is
\begin{align*}
\Psi_{\mathcal{A}}^2 \equiv\ \{& \{\Xi_1,\Xi_2\}, \{\Xi_1, \Xi_4\},  \{\Xi_1, \Xi_2+a_1\Xi_4\}, \{\Xi_3, \Xi_4\}\}.
\end{align*}

The graph of the 2D-optimal system computed by \symbolie is shown in Figure~\ref{fig:graph2D_10}.
\begin{figure}
\centering
\includegraphics[width=.4\textwidth]{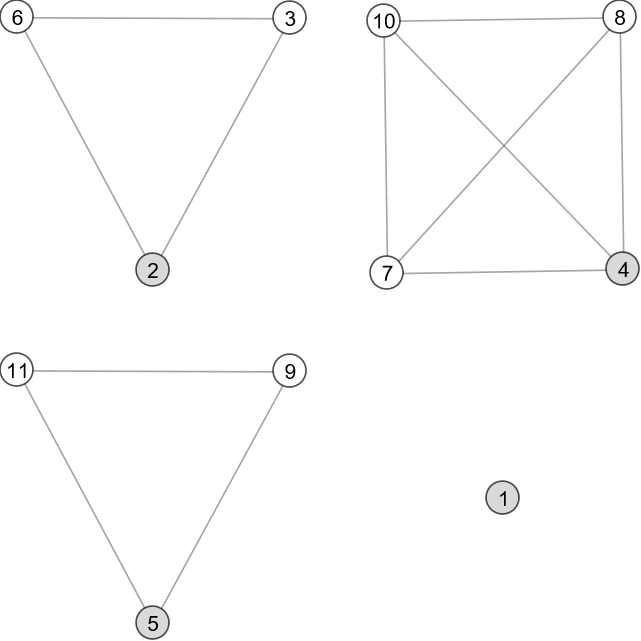}
\label{fig:graph2D_10}
\caption{Two-dimensional optimal system for the Lie algebra $A_{3,7}^a \oplus A_1$ computed by \symbolie. \footnotesize\label{fig:graph4D10} $1\to \{\Xi_1,\Xi_2\}$, $2\to \{\Xi_1,\Xi_4\}$, $3\to \{\Xi_2,\Xi_4\}$, $4\to \{\Xi_3,\Xi_4\}$, $5\to \{\Xi_1,\Xi_2+a_1 \Xi_4\}$, $6\to \{\Xi_1+a_1 \Xi_2,\Xi_4\}$, $7\to \{\Xi_1+a_1 \Xi_3,\Xi_4\}$, $8\to \{\Xi_2+a_1 \Xi_3,\Xi_4\}$, $9\to \{\Xi_1+a_1 \Xi_4,\Xi_2\}$, $10\to \{\Xi_1+a_1 \Xi_2+a_2 \Xi_3,\Xi_4\}$, $11\to \{\Xi_1+a_1 \Xi_4,\Xi_2+a_2 \Xi_4\}$}
\end{figure}

Applying the inner automorphism $A_3$ to $ \{\Xi_1, \Xi_2+a_1\Xi_4\}$, we obtain
\[
\{\exp(at_3)(\cos t_3\Xi_1 - \sin t_3\Xi_2), \exp(at_3)(\sin t_3\Xi_1 +\cos t_3\Xi_2 + a_1\Xi_4) \},
\]
and, choosing the parameter $t=\pi/2$, we have
\[
\{-\exp(a\pi/2)\Xi_2, \exp(a\pi/2)\Xi_1 + a_1\Xi_4 \}.
\]
Finally, rescaling by the factor $-\exp(a\pi/2)$ we obtain the family of subalgebras 
$\{\Xi_1+x\Xi_4, \Xi_2\}$.
\end{algebra}

\begin{algebra}[$\mathbf{A_{4,5}^{a,b}}$]
Let $\mathcal{L}_4$ be the Lie algebra spanned by $\{\Xi_1,\Xi_2,\Xi_3,\Xi_4\}$ , whose only nonvanishing commutators are
\[
[\Xi_1,\Xi_4]=\Xi_1,\quad [\Xi_2,\Xi_4]=a\Xi_2,\quad [\Xi_3,\Xi_4]=b\Xi_3,
\]
where $-1\le a<b<1,$ and $ab\neq 0$. The matrices associated to inner automorphisms $\exp(t_k \ad_{\Xi_k})$, $k=1,\ldots,4$, are
\begin{gather*}
A_1 =
\begin{pmatrix}
1 & 0 & 0 & -t_1\\
0 & 1 & 0 & 0\\
0 & 0 & 1 & 0\\
0 & 0 & 0 & 1
\end{pmatrix},\;
A_2 =
\begin{pmatrix}
1 & 0 & 0 & 0\\
0 & 1 & 0 & -a t_2\\
0 & 0 & 1 & 0\\
0 & 0 & 0 & 1
\end{pmatrix},\;
A_3 =
\begin{pmatrix}
1 & 0 & 0 & 0\\
0 & 1 & 0 & 0\\
0 & 0 & 1 & -b t_3\\
0 & 0 & 0 & 1
\end{pmatrix},\\
A_4 =
\begin{pmatrix}
\exp(t_4) & 0 & 0 & 0\\
0 & \exp(a t_4) & 0 & 0\\
0 & 0 & \exp(b t_4) & 0\\
0 & 0 & 0 & 1
\end{pmatrix}.
\end{gather*}

The one-dimensional optimal system computed by \symbolie is
\begin{align*}
\Psi^1_{\mathcal{A}} = \Theta_A^1 \cup \{\Xi_1+\alpha_1\Xi_2\},
\end{align*}
with $\alpha_1=\pm 1$ and its graph is shown in Figure~\ref{fig:graph4D18}.

\begin{figure}[h]
\centering
\includegraphics[width=.4\textwidth]{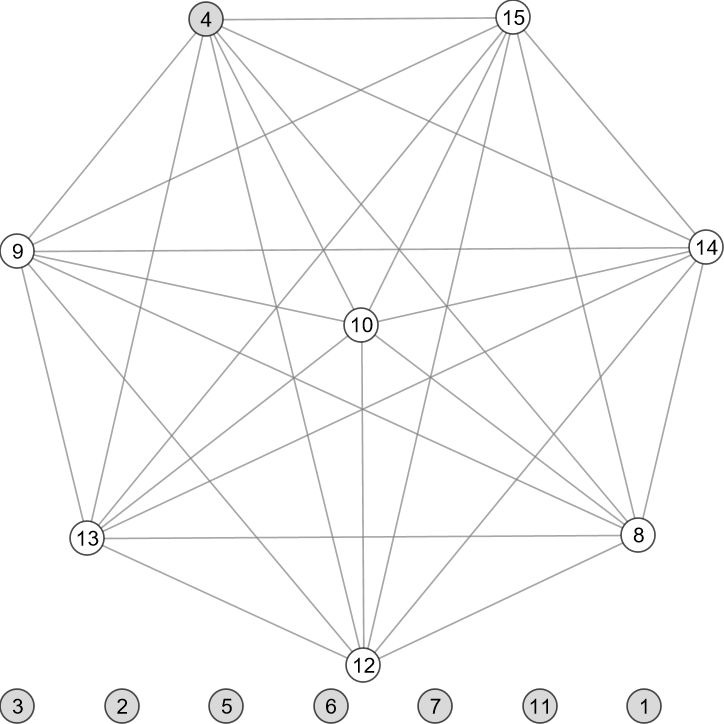}
\label{fig:graph4D18}
\caption{One-dimensional optimal system for the Lie algebra $A_{4,5}^{a,b}$ computed by \symbolie. \footnotesize $1\to \{\Xi_1\}$, $2\to \{\Xi_2\}$,$3\to \{\Xi_3\}$,$4\to \{\Xi_4\}$,$5\to \{\Xi_1+a_1\Xi_2\}$,$6\to \{\Xi_1+a_1 \Xi_3\}$,$7\to \{\Xi_2+a_1 \Xi_3\}$,$8\to \{\Xi_1+a_1 \Xi_4\}$,$9\to \{\Xi_2+a_1 \Xi_4\}$,$10\to \{\Xi_3+a_1 \Xi_4\}$,$11\to \{\Xi_1+a_1 \Xi_2+a_2 \Xi_3\}$,$12\to \{\Xi_1+a_1 \Xi_2+a_2 \Xi_4\}$,$13\to \{\Xi_1+a_1 \Xi_3+a_2 \Xi_4\}$,$14\to \{\Xi_2+a_1 \Xi_3+a_2 \Xi_4\}$,$15\to \{\Xi_1+a_1 \Xi_2+a_2 \Xi_3+a_3 \Xi_4\}$}
\end{figure}

In \cite[Table II]{PW} there is all representatives with the exception of $\{\Xi_1+\alpha_1\Xi_2\}$. We observe that such family of subalgebras is invariant with respect to the action of the inner automorphisms, 
so $\{\Xi_1+\alpha_1\Xi_2\}$ has to belong to the optimal system.
\end{algebra}

\begin{algebra}[$\mathbf{A_{4,6}^{a,b}}$]
Let $\mathcal{L}_4$ be the Lie algebra spanned by $\{\Xi_1,\Xi_2,\Xi_3,\Xi_4\}$, whose nonzero commutators are
\[
[\Xi_1,\Xi_4]=a\Xi_1,\quad [\Xi_2,\Xi_4]=b\Xi_2-\Xi_3,\quad [\Xi_3,\Xi_4]=\Xi_2+b\Xi_3,
\]
where $a\neq 0$ and $b\geq 0$. The matrices associated to inner automorphisms $\exp(t_k \ad_{\Xi_k})$, $k=1,\ldots,4$, are
\begin{gather*}
A_1 =
\begin{pmatrix}
1 & 0 & 0 & -at_1\\
0 & 1 & 0 & 0\\
0 & 0 & 1 & 0\\
0 & 0 & 0 & 1
\end{pmatrix},\;
A_2 =
\begin{pmatrix}
1 & 0 & 0 & 0\\
0 & 1 & 0 & -b t_2\\
0 & 0 & 1 & t\\
0 & 0 & 0 & 1
\end{pmatrix},\;
A_3 =
\begin{pmatrix}
1 & 0 & 0 & 0\\
0 & 1 & 0 & -t_3\\
0 & 0 & 1 & -bt_3\\
0 & 0 & 0 & 1
\end{pmatrix},\\
A_4 =
\begin{pmatrix}
\exp(a t_4) & 0 & 0 & 0\\
0 & \exp(b t_4)\cos (t_4) & \exp(b t_4)\sin (t_4) & 0\\
0 & -\exp(b t_4)\sin (t_4) & \exp(b t_4)\cos (t_4) & 0\\
0 & 0 & 0 & 1
\end{pmatrix}.
\end{gather*}

In \cite{PW}, the authors listed the one-dimensional optimal system
\[
\Theta_{\mathcal{A}}^1 \equiv\  \{ \{\Xi_1\},\{\Xi_3\}, \{\Xi_1+x\Xi_3\}, \{\Xi_4\} \},
\]
with $x>0$. Using \symbolie, we obtain the one-dimensional optimal system of subalgebras:
\[
\Psi_{\mathcal{A}}^1 \equiv\  \{ \{\Xi_1\},\{\Xi_2\}, \{\Xi_1+a_1\Xi_2\}, \{\Xi_4\} \},
\]
with $a_1\in \mathbb{R}^\star$. The corresponding graph is illustrated in Figure~\ref{fig:graph4D22}.

\begin{figure}[h]
\centering
\includegraphics[width=.4\textwidth]{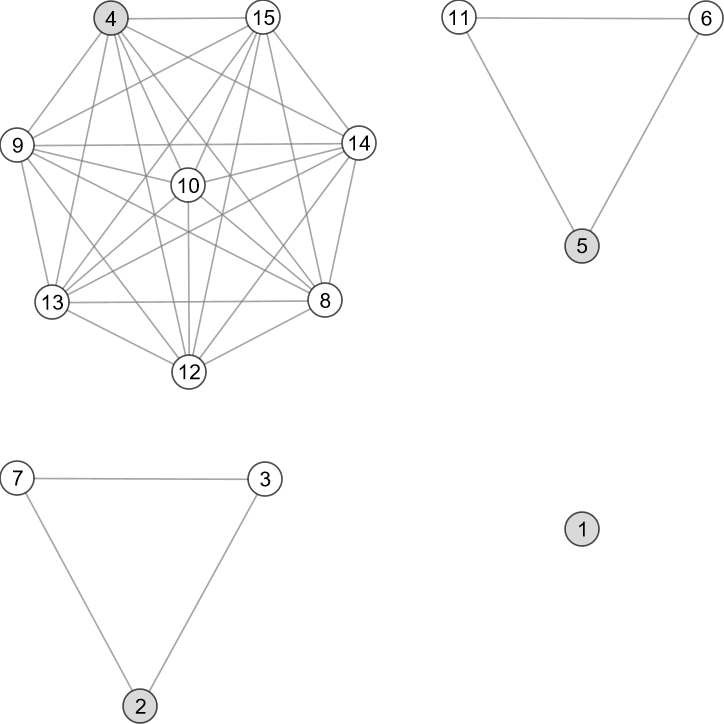}
\label{fig:graph4D22}
\caption{One-dimensional optimal system for the Lie algebra $A_{4,6}^{a,b}$ computed by \symbolie. \footnotesize $1\to \{\Xi_1\}$, $2\to \{\Xi_2\}$,$3\to \{\Xi_3\}$,$4\to \{\Xi_4\}$,$5\to \{\Xi_1+a_1\Xi_2\}$,$6\to \{\Xi_1+a_1 \Xi_3\}$,$7\to \{\Xi_2+a_1 \Xi_3\}$,$8\to \{\Xi_1+a_1 \Xi_4\}$,$9\to \{\Xi_2+a_1 \Xi_4\}$,$10\to \{\Xi_3+a_1 \Xi_4\}$,$11\to \{\Xi_1+a_1 \Xi_2+a_2 \Xi_3\}$,$12\to \{\Xi_1+a_1 \Xi_2+a_2 \Xi_4\}$,$13\to \{\Xi_1+a_1 \Xi_3+a_2 \Xi_4\}$,$14\to \{\Xi_2+a_1 \Xi_3+a_2 \Xi_4\}$,$15\to \{\Xi_1+a_1 \Xi_2+a_2 \Xi_3+a_3 \Xi_4\}$}
\end{figure}

Let us show that there is an inner automorphism that maps $\{\Xi_2\}$ into $\{\Xi_3\}$, and then $\{\Xi_1+a_1\Xi_2\}$ into $\{\Xi_1+x\Xi_3\}$:
\[
(0,1,0,0)^T \stackrel{A_4}{\mapsto}  (0,\exp(bt)\cos (t), - \exp(bt)\sin (t), 0)= (0,0,-\exp(b\pi/2),0) \quad \text{with}\ t = \frac{\pi}{2},
\]
from which it follows that the subalgebra $\{\Xi_2\}$ is equivalent to $\{\Xi_3\}$.

Similarly, we have
\[
(1,x,0,0)\stackrel{A_4}{\mapsto}  (\exp(a\pi/2),0,-\exp(b\pi/2)x,0),\quad \text{with}\ t = \frac{\pi}{2},
\]
and, since $\{\exp(a\pi/2)\Xi_1-\exp(b\pi/2)x\Xi_3\} = \{\Xi_1 + a_1\Xi_3\}$, with $a_1 = -\exp(\pi(b-a)/2)x$ arbitrary in $\mathbb{R}^\star$, the claim is proved.

Moreover, the two-dimensional optimal systems $\Theta_{\mathcal{A}}^2$ and $\Psi_{\mathcal{A}}^2$ are 
\[
\Theta_{\mathcal{A}}^2 \equiv\ \{
\{\Xi_1,\Xi_2\}, \{\Xi_2, \Xi_3\}, \{\Xi_1,\Xi_4\}, \{\Xi_1+x\Xi_3,\Xi_2\}\},\quad x > 0,
\]
and
\[
\Psi^2_{\mathcal{A}} \equiv\ \{
\{\Xi_1,\Xi_2\}, \{\Xi_2, \Xi_3\}, \{\Xi_1,\Xi_4\}, \{\Xi_1+a_1\Xi_2,\Xi_3\}\},\quad a_1 \in \mathbb{R}^\star.
\]
The optimal systems coincide, once we note from the graph below that the $3$-families $\{\Xi_1+x\Xi_3,\Xi_2\}$ and  $\{\Xi_1+a_1\Xi_2,\Xi_3\}$ belong to the same connected component.

\begin{figure}[h]
\centering
\includegraphics[width=.4\textwidth]{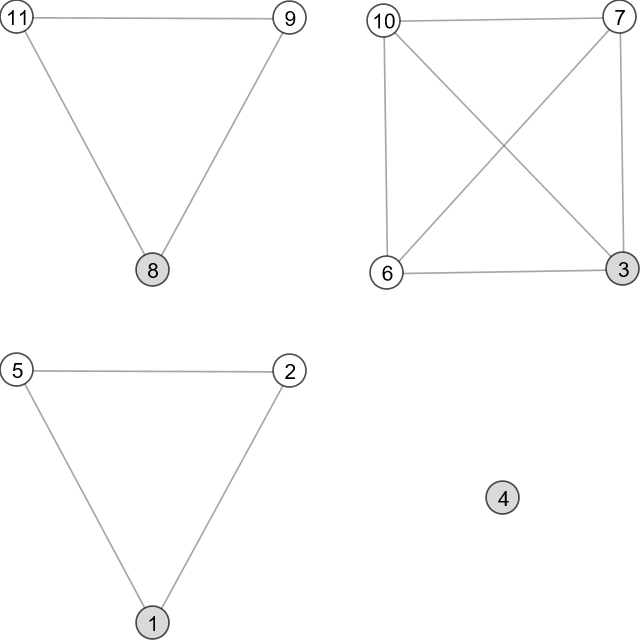}
\caption{Two-dimensional optimal system for the Lie algebra $A_{4,6}^{a,b}$ computed by \symbolie.\footnotesize\label{fig:graph4D22_2} $1\to \{\Xi_1,\Xi_2\}$, $2\to \{\Xi_1,\Xi_3\}$, $3\to \{\Xi_1,\Xi_4\}$, $4\to \{\Xi_2,\Xi_3\}$, $5\to \{\Xi_1,a_1 \Xi_2+a_2 \Xi_3\}$, $6\to \{\Xi_1,\Xi_2+a_1 \Xi_4\}$, $7\to \{\Xi_1,\Xi_3+a_1 \Xi_4\}$, $8\to \{\Xi_1+a_1 \Xi_2,\Xi_3\}$, $9\to \{\Xi_1+a_1 \Xi_3,\Xi_2\}$, $10\to \{\Xi_1,\Xi_2+a_1 \Xi_3+a_2 \Xi_4\}$, $11\to \{\Xi_1+a_1 \Xi_3,\Xi_2+a_2 \Xi_3\}$}
\end{figure}

In particular, the two families under consideration are labeled with $8$ and $9$, and we can see from graph that are equivalent. Indeed, using the inner automorphism $A_4$ and setting $t = \pi/2$, as we have already seen, we obtain the equivalence.
\end{algebra}

In Table \ref{tab:optimal4d}, the optimal systems of four-dimensional Lie algebras, as found by \symbolie, are listed. Also in this case the computation requires a few minutes (the file with the computation can be found at the url \url{https://mat521.unime.it/oliveri}).

\smallsize
\begin{center}
\setlength{\LTcapwidth}{0.98\textwidth}
\begin{longtable}{p{1.2cm}p{2.7cm}p{3cm}p{3cm}p{3cm}}
\caption{\label{tab:optimal4d}Optimal systems of real Lie algebras of dimension 4. The constants $a_1,a_2,a_3,a_4\in \mathbb{R}$ are nonvanishing, and $\alpha_1=\pm 1$. 
\texttt{SymboLie} completes the computation in a few minutes.} \\
\toprule
\textbf{Algebra} & \begin{minipage}{2.7cm}\textbf{Nonzero\\ Lie brackets} \end{minipage}&
\textbf{1D Optimal System} & \textbf{2D Optimal System} &
\textbf{3D Optimal System} \\
\midrule
\endfirsthead

\caption[]{(continued)}\\
\toprule
\textbf{Algebra} & \begin{minipage}{2.7cm}\textbf{Nonzero\\ Lie brackets}\end{minipage} &
\textbf{1D Optimal System} & \textbf{2D Optimal System} &
\textbf{3D Optimal System} \\
\midrule
\endhead

\midrule
\multicolumn{5}{r}{\textit{Continued}}\\
\midrule
\endfoot

\bottomrule
\endlastfoot

$4A_{1}$ & & All distinct one-dimensional Lie subalgebras (15) & 
All distinct two-dimensional Lie subalgebras (35) &
All distinct three-dimensional Lie subalgebras (15) \\
\noalign{\smallskip}\hline\noalign{\smallskip}
$A_2\oplus 2A_1$& 
\begin{minipage}{2.7cm}
$[\Xi_1,\Xi_2]=\Xi_2$
\end{minipage}&  
\begin{minipage}{3cm}
$\{\Xi_1\}$, $\{\Xi_2\}$, \\
$\{\Xi_3\}$, $\{\Xi_4\}$, \\
$\{\Xi_1+a_1\Xi_3\}$,\\
$\{\Xi_2+\alpha_1\Xi_3\}$, \\
$\{\Xi_1+a_1\Xi_4\}$, \\
$\{\Xi_2+\alpha_1\Xi_4\}$,\\
$\{\Xi_3+a_1\Xi_4\}$,\\
$\{\Xi_1+a_1\Xi_3+a_2\Xi_4\}$,\\
$\{\Xi_2+\alpha_1\Xi_3+a_1\Xi_4\}$
\end{minipage} &  
\begin{minipage}{3cm}
$\{\Xi_1,\Xi_2\}$, $\{\Xi_1,\Xi_3\}$, \\
$\{\Xi_1,\Xi_4\}$, $\{\Xi_2,\Xi_3\}$, \\
$\{\Xi_2,\Xi_4\}$, $\{\Xi_3,\Xi_4\}$, \\
$\{\Xi_1,\Xi_3+a_1\Xi_4\}$, \\
$\{\Xi_2,\Xi_3+a_1\Xi_4\}$, \\
$\{\Xi_1+a_1\Xi_3,\Xi_2\}$, \\
$\{\Xi_1+a_1\Xi_3,\Xi_4\}$, \\
$\{\Xi_2+\alpha_1\Xi_3,\Xi_4\}$, \\
$\{\Xi_1+a_1\Xi_4,\Xi_2\}$, \\
$\{\Xi_1+a_1\Xi_4,\Xi_3\}$, \\
$\{\Xi_2+\alpha_1\Xi_4,\Xi_3\}$, \\
$\{\Xi_1+a_1\Xi_4,\Xi_3+a_2\Xi_4\}$, \\
$\{\Xi_2+\alpha_1\Xi_4,\Xi_3+a_1\Xi_4\}$, \\
$\{\Xi_1+a_1\Xi_3+a_2\Xi_4,\Xi_2\}$
\end{minipage}   &
\begin{minipage}{3cm}
$\{\Xi_1,\Xi_2,\Xi_3\}$,\\
$\{\Xi_1,\Xi_2,\Xi_4\}$,\\
$\{\Xi_1,\Xi_3,\Xi_4\}$,\\
$\{\Xi_2,\Xi_3,\Xi_4\}$,\\
$\{\Xi_1,\Xi_2,\Xi_3+a_1\Xi_4\}$,\\
$\{\Xi_1+a_1\Xi_3,\Xi_2,\Xi_4\}$,\\
$\{\Xi_1+a_1\Xi_4,\Xi_2,\Xi_3\}$,\\
$\{\Xi_1+a_1\Xi_4,\Xi_2,\Xi_3+$\\
$\phantom{\{}a_2\Xi_4\}$
\end{minipage}\\
\noalign{\smallskip}\hline\noalign{\smallskip}
 $2A_2$ &
 \begin{minipage}{2.7cm}
$[\Xi_1,\Xi_2]=\Xi_2$\\
$[\Xi_3,\Xi_4]=\Xi_4$
\end{minipage}&  
\begin{minipage}{3cm}
$\{\Xi_1\}$, $\{\Xi_2\}$, \\
$\{\Xi_3\}$, $\{\Xi_4\}$, \\
$\{\Xi_1+a_1 \Xi_3\}$,\\ 
$\{\Xi_2+\alpha_1 \Xi_3\}$, \\
$\{\Xi_1+\alpha_1 \Xi_4\}$,\\
$\{\Xi_2+\alpha_1 \Xi_4\}$
\end{minipage} &  
\begin{minipage}{3cm}
$\{\Xi_1,\Xi_2\}$, $\{\Xi_1,\Xi_3\}$, \\
$\{\Xi_1,\Xi_4\}$, $\{\Xi_2,\Xi_3\}$, \\
$\{\Xi_2,\Xi_4\}$, $\{\Xi_3,\Xi_4\}$, \\
$\{\Xi_1+a_1\Xi_3,\Xi_2\}$, \\
$\{\Xi_1+a_1\Xi_3,\Xi_4\}$, \\
$\{\Xi_2+\alpha_1\Xi_3,\Xi_4\}$, \\
$\{\Xi_1+\alpha_1\Xi_4,\Xi_2\}$
\end{minipage}   &
\begin{minipage}{3cm}
$\{\Xi_1,\Xi_2,\Xi_3\}$,\\
$\{\Xi_1,\Xi_2,\Xi_4\}$,\\
$\{\Xi_1,\Xi_3,\Xi_4\}$,\\
$\{\Xi_2,\Xi_3,\Xi_4\}$,\\
$\{\Xi_1+a_1\Xi_3,\Xi_2,\Xi_4\}$
\end{minipage}\\
\noalign{\smallskip}\hline\noalign{\smallskip}
$A_{3,1}\oplus A_1$ & 
\begin{minipage}{2.7cm}
$[\Xi_2,\Xi_3]=\Xi_1$
\end{minipage}&
\begin{minipage}{3cm}
$\{\Xi_1\}$, $\{\Xi_2\}$, \\
$\{\Xi_3\}$, $\{\Xi_4\}$, \\
$\{\Xi_2+a_1\Xi_3\}$,\\
$\{\Xi_1+a_1\Xi_4\}$, \\
$\{\Xi_2+a_1\Xi_4\}$, \\
$\{\Xi_3+a_1\Xi_4\}$, \\
$\{\Xi_2+a_1\Xi_3+a_2\Xi_4\}$
\end{minipage} &  
\begin{minipage}{3cm}
$\{\Xi_1,\Xi_2\}$, $\{\Xi_1,\Xi_3\}$,\\
$\{\Xi_1,\Xi_4\}$, $\{\Xi_2,\Xi_4\}$,\\ 
$\{\Xi_3,\Xi_4\}$,\\
$\{\Xi_1,\Xi_2+a_ 1 \Xi_3\}$,\\
$\{\Xi_1,\Xi_2+a_1 \Xi_4\}$,\\ 
$\{\Xi_1,\Xi_3+a_1 \Xi_4\}$,\\
$\{\Xi_2+a_1 \Xi_3,\Xi_4\}$,\\
$\{\Xi_1+a_1 \Xi_4,\Xi_2\}$,\\
$\{\Xi_1+a_1 \Xi_4,\Xi_3\}$,\\
$\{\Xi_1,\Xi_2+a_1 \Xi_3+a_2 \Xi_4\}$,\\
$\{\Xi_1+a_1 \Xi_4,\Xi_2+a_2 \Xi_3\}$
\end{minipage}   &
\begin{minipage}{3cm}
$\{\Xi_1,\Xi_2,\Xi_3\}$, $\{\Xi_1,\Xi_2,\Xi_4\}$, \\
$\{\Xi_1,\Xi_3,\Xi_4\}$, \\
$\{\Xi_1,\Xi_2,\Xi_3+a_1 \Xi_4\}$,\\
$\{\Xi_1,\Xi_2+a_1 \Xi_3,\Xi_4\}$,\\
$\{\Xi_1,\Xi_2+a_1 \Xi_4,\Xi_3\}$, \\
$\{\Xi_1,\Xi_2+a_1 \Xi_4,\Xi_3+$\\
$\phantom{\{}a_2 \Xi_4\}$
\end{minipage}\\
\noalign{\smallskip}\hline\noalign{\smallskip}
$A_{3,2}\oplus A_1$ & 
\begin{minipage}{2.7cm}
{\footnotesize $[\Xi_1,\Xi_3]=\Xi_1$}\\
{\footnotesize $[\Xi_2,\Xi_3]=\Xi_1+\Xi_2$}
\end{minipage}&
\begin{minipage}{3cm}
$\{\Xi_1\}$, $\{\Xi_2\}$, \\
$\{\Xi_3\}$, $\{\Xi_4\}$, \\
$\{\Xi_1+\alpha_1 \Xi_4\}$,\\ 
$\{\Xi_2+a_1 \Xi_4\}$,\\
$\{\Xi_3+a_1 \Xi_4\}$
\end{minipage} &  
\begin{minipage}{3cm}
$\{\Xi_1,\Xi_2\}$, $\{\Xi_1,\Xi_3\}$,\\
$\{\Xi_1,\Xi_4\}$, $\{\Xi_2,\Xi_4\}$,\\
$\{\Xi_3,\Xi_4\}$,\\
$\{\Xi_1,\Xi_2+\alpha_1 \Xi_4\}$,\\
$\{\Xi_1,\Xi_3+a_1 \Xi_4\}$,\\
$\{\Xi_1+a_1 \Xi_4,\Xi_2\}$
\end{minipage}   &
\begin{minipage}{3cm}
$\{\Xi_1,\Xi_2,\Xi_3\}$, $\{\Xi_1,\Xi_2,\Xi_4\}$,\\
$\{\Xi_1,\Xi_3,\Xi_4\}$,\\
$\{\Xi_1,\Xi_2,\Xi_3+a_1 \Xi_4\}$
\end{minipage}\\
\noalign{\smallskip}\hline\noalign{\smallskip}
$A_{3,3}\oplus A_1$ & 
\begin{minipage}{2.7cm}
$[\Xi_1,\Xi_3]=\Xi_1$\\
$[\Xi_2,\Xi_3]=\Xi_2$
\end{minipage}&
\begin{minipage}{3cm}
$\{\Xi_1\}$, $\{\Xi_2\}$, \\
$\{\Xi_3\}$, $\{\Xi_4\}$,\\
$\{\Xi_1+a_1 \Xi_2\}$,\\
$\{\Xi_1+\alpha_1 \Xi_4\}$,\\
$\{\Xi_2+\alpha_1 \Xi_4\}$,\\
$\{\Xi_3+a_1 \Xi_4\}$,\\
$\{\Xi_1+a_1 \Xi_2+\alpha_1 \Xi_4\}$
\end{minipage} &  
\begin{minipage}{3cm}
$\{\Xi_1,\Xi_2\}$, $\{\Xi_1,\Xi_3\}$,\\
$\{\Xi_1,\Xi_4\}$, $\{\Xi_2,\Xi_3\}$,\\
$\{\Xi_2,\Xi_4\}$, $\{\Xi_3,\Xi_4\}$,\\
$\{\Xi_1,\Xi_2+\alpha_1 \Xi_4\}$,\\
$\{\Xi_1,\Xi_3+a_1 \Xi_4\}$,\\
$\{\Xi_2,\Xi_3+a_1 \Xi_4\}$,\\
$\{\Xi_1+a_1 \Xi_2,\Xi_3\}$,\\
$\{\Xi_1+a_1 \Xi_2,\Xi_4\}$,\\
$\{\Xi_1+\alpha_1 \Xi_4,\Xi_2\}$,\\
$\{\Xi_1+a_1 \Xi_2,\Xi_3+a_2 \Xi_4\}$,\\
$\{\Xi_1+\alpha_1 \Xi_4,\Xi_2+a_1 \Xi_4\}$
\end{minipage}&
\begin{minipage}{3cm}
$\{\Xi_1,\Xi_2,\Xi_3\}$, $\{\Xi_1,\Xi_2,\Xi_4\}$,\\
$\{\Xi_1,\Xi_3,\Xi_4\}$, $\{\Xi_2,\Xi_3,\Xi_4\}$,\\
$\{\Xi_1,\Xi_2,\Xi_3+a_1 \Xi_4\}$,\\
$\{\Xi_1+a_1 \Xi_2,\Xi_3,\Xi_4\}$
\end{minipage}\\
\noalign{\smallskip}\hline\noalign{\smallskip}
$A_{3,4}\oplus A_1$ & 
\begin{minipage}{2.7cm}
$[\Xi_1,\Xi_3]=\Xi_1$\\
$[\Xi_2,\Xi_3]=-\Xi_2$
\end{minipage}&
\begin{minipage}{3cm}
$\{\Xi_1\}$, $\{\Xi_2\}$, \\
$\{\Xi_3\}$, $\{\Xi_4\}$,\\
$\{\Xi_1+\alpha_1 \Xi_2\}$,\\
$\{\Xi_1+\alpha_1 \Xi_4\}$,\\
$\{\Xi_2+\alpha_1 \Xi_4\}$, \\
$\{\Xi_3+a_1 \Xi_4\}$,\\
$\{\Xi_1+\alpha_1 \Xi_2+a_1 \Xi_4\}$
\end{minipage} &  
\begin{minipage}{3cm}
$\{\Xi_1,\Xi_2\}$, $\{\Xi_1,\Xi_3\}$,\\
$\{\Xi_1,\Xi_4\}$, $\{\Xi_2,\Xi_3\}$,\\
$\{\Xi_2,\Xi_4\}$, $\{\Xi_3,\Xi_4\}$,\\
$\{\Xi_1,\Xi_2+\alpha_1 \Xi_4\}$,\\
$\{\Xi_1,\Xi_3+a_1 \Xi_4\}$,\\
$\{\Xi_2,\Xi_3+a_1 \Xi_4\}$,\\
$\{\Xi_1+\alpha_1 \Xi_2,\Xi_4\}$,\\
$\{\Xi_1+\alpha_1 \Xi_4,\Xi_2\}$,\\
$\{\Xi_1+\alpha_1 \Xi_4,\Xi_2+a_1 \Xi_4\}$
\end{minipage}&
\begin{minipage}{3cm}
$\{\Xi_1,\Xi_2,\Xi_3\}$, $\{\Xi_1,\Xi_2,\Xi_4\}$,\\
$\{\Xi_1,\Xi_3,\Xi_4\}$, $\{\Xi_2,\Xi_3,\Xi_4\}$,\\
$\{\Xi_1,\Xi_2,\Xi_3+a_1 \Xi_4\}$
\end{minipage}\\
\noalign{\smallskip}\hline\noalign{\smallskip}
$A_{3,5}^a\oplus A_1$ & 
\begin{minipage}{2.7cm}
$[\Xi_1,\Xi_3]=\Xi_1$\\
$[\Xi_2,\Xi_3]=a\Xi_2$\\
$0<|a|<1$
\end{minipage}&
\begin{minipage}{3cm}
$\{\Xi_1\}$, $\{\Xi_2\}$, \\
$\{\Xi_3\}$, $\{\Xi_4\}$,\\
$\{\Xi_1+\alpha_1 \Xi_2\}$, \\
$\{\Xi_1+\alpha_1 \Xi_4\}$,\\
$\{\Xi_2+\alpha_1 \Xi_4\}$,\\
$\{\Xi_3+a_1 \Xi_4\}$,\\
$\{\Xi_1+\alpha_1 \Xi_2+a_1 \Xi_4\}$
\end{minipage} &  
\begin{minipage}{3cm}
$\{\Xi_1,\Xi_2\}$, $\{\Xi_1,\Xi_3\}$,\\
$\{\Xi_1,\Xi_4\}$, $\{\Xi_2,\Xi_3\}$,\\
$\{\Xi_2,\Xi_4\}$, $\{\Xi_3,\Xi_4\}$,\\
$\{\Xi_1,\Xi_2+\alpha_1 \Xi_4\}$,\\
$\{\Xi_1,\Xi_3+a_1 \Xi_4\}$,\\
$\{\Xi_2,\Xi_3+a_1 \Xi_4\}$,\\
$\{\Xi_1+\alpha_1 \Xi_2,\Xi_4\}$,\\
$\{\Xi_1+\alpha_1 \Xi_4,\Xi_2\}$,\\
$\{\Xi_1+\alpha_1 \Xi_4,\Xi_2+a_1 \Xi_4\}$
\end{minipage}&
\begin{minipage}{3cm}
$\{\Xi_1,\Xi_2,\Xi_3\}$, $\{\Xi_1,\Xi_2,\Xi_4\}$,\\
$\{\Xi_1,\Xi_3,\Xi_4\}$, $\{\Xi_2,\Xi_3,\Xi_4\}$,\\
$\{\Xi_1,\Xi_2,\Xi_3+a_1 \Xi_4\}$
\end{minipage}\\
\noalign{\smallskip}\hline\noalign{\smallskip}
$A_{3,6}\oplus A_1$ & 
\begin{minipage}{2.7cm}
$[\Xi_1,\Xi_3]=-\Xi_2$\\
$[\Xi_2,\Xi_3]=\Xi_1$
\end{minipage}&
\begin{minipage}{3cm}
$\{\Xi_1\}$, $\{\Xi_3\}$, $\{\Xi_4\}$,\\
$\{\Xi_1+a_1 \Xi_4\}$, \\
$\{\Xi_3+a_1 \Xi_4\}$
\end{minipage} &  
\begin{minipage}{3cm}
$\{\Xi_1,\Xi_2\}$, $\{\Xi_1,\Xi_4\}$,\\
$\{\Xi_3,\Xi_4\}$,\\
$\{\Xi_1,\Xi_2+a_1 \Xi_4\}$
\end{minipage}&
\begin{minipage}{3cm}
$\{\Xi_1,\Xi_2,\Xi_3\}$, $\{\Xi_1,\Xi_2,\Xi_4\}$,\\
$\{\Xi_1,\Xi_2,\Xi_3+a_1 \Xi_4\}$
\end{minipage}\\
\noalign{\smallskip}\hline\noalign{\smallskip}
$A_{3,7}^a\oplus A_1$ &
\begin{minipage}{2.7cm}
$[\Xi_1,\Xi_3]=a\Xi_1-\Xi_2$\\
$[\Xi_2,\Xi_3]=\Xi_1+a\Xi_2$\\
$a > 0$
\end{minipage}&
\begin{minipage}{3cm}
$\{\Xi_1\}$, $\{\Xi_3\}$, $\{\Xi_4\}$,\\
$\{\Xi_1+a_1 \Xi_4\}$,\\
$\{\Xi_3+a_1 \Xi_4\}$
\end{minipage} &  
\begin{minipage}{3cm}
$\{\Xi_1,\Xi_2\}$, $\{\Xi_1,\Xi_4\}$,\\
$\{\Xi_3,\Xi_4\}$,\\
$\{\Xi_1,\Xi_2+a_1 \Xi_4\}$
\end{minipage}   &
\begin{minipage}{3cm}
$\{\Xi_1,\Xi_2,\Xi_3\}$, $\{\Xi_1,\Xi_2,\Xi_4\}$,\\
$\{\Xi_1,\Xi_2,\Xi_3+a_1 \Xi_4\}$
\end{minipage}\\
\noalign{\smallskip}\hline\noalign{\smallskip}
$A_{3,8}\oplus A_1$ &
\begin{minipage}{2.7cm}
$[\Xi_1,\Xi_3]=-2\Xi_2$\\
$[\Xi_1,\Xi_2]=\Xi_1$\\
$[\Xi_2,\Xi_3]=\Xi_3$
\end{minipage}&  
\begin{minipage}{3cm}
$\{\Xi_1\}$, $\{\Xi_2\}$, $\{\Xi_4\}$,\\
$\{\Xi_1+\alpha_1 \Xi_3\}$,\\
$\{\Xi_1+\alpha_1 \Xi_4\}$,\\
$\{\Xi_2+a_1 \Xi_4\}$,\\
$\{\Xi_1+\alpha_1 \Xi_3+a_1 \Xi_4\}$
\end{minipage} &  
\begin{minipage}{3cm}
$\{\Xi_1,\Xi_2\}$, $\{\Xi_1,\Xi_4\}$,\\
$\{\Xi_2,\Xi_4\}$,\\
$\{\Xi_1,\Xi_2+a_1 \Xi_4\}$,\\
$\{\Xi_1+\alpha_1 \Xi_3,\Xi_4\}$
\end{minipage} &
\begin{minipage}{3cm}
$\{\Xi_1,\Xi_2,\Xi_3\}$, $\{\Xi_1,\Xi_2,\Xi_4\}$
\end{minipage}\\
\noalign{\smallskip}\hline\noalign{\smallskip}
$A_{3,9}\oplus A_1$ &
\begin{minipage}{2.7cm}
$[\Xi_1,\Xi_2]=\Xi_3$\\
$[\Xi_2,\Xi_3]=\Xi_1$\\
$[\Xi_1, \Xi_3]=-\Xi_2$
\end{minipage}&  
\begin{minipage}{3cm}
$\{\Xi_1\}$, $\{\Xi_4\}$, \\
$\{\Xi_1+a_1 \Xi_4\}$
\end{minipage} &  
\begin{minipage}{3cm}
$\{\Xi_1,\Xi_4\}$
\end{minipage}   &
\begin{minipage}{3cm}
$\{\Xi_1,\Xi_2,\Xi_3\}$
\end{minipage}\\
\noalign{\smallskip}\hline\noalign{\smallskip}
$A_{4,1}$ & 
\begin{minipage}{2.7cm}
$[\Xi_2,\Xi_4]=\Xi_1$\\
$[\Xi_3,\Xi_4]=\Xi_2$
\end{minipage}&  
\begin{minipage}{3cm}
$\{\Xi_1\}$, $\{\Xi_2\}$, \\
$\{\Xi_3\}$, $\{\Xi_4\}$,\\
$\{\Xi_1+a_1 \Xi_3\}$, \\
$\{\Xi_3+a_1 \Xi_4\}$
\end{minipage} &  
\begin{minipage}{3cm}
$\{\Xi_1,\Xi_2\}$, $\{\Xi_1,\Xi_3\}$,\\
$\{\Xi_1,\Xi_4\}$, $\{\Xi_2,\Xi_3\}$,\\
$\{\Xi_1,\Xi_3+a_1 \Xi_4\}$,\\
$\{\Xi_1+a_1 \Xi_3,\Xi_2\}$
\end{minipage}   &
\begin{minipage}{3cm}
$\{\Xi_1,\Xi_2,\Xi_3\}$, $\{\Xi_1,\Xi_2,\Xi_4\}$,\\
$\{\Xi_1,\Xi_2,\Xi_3+a_1 \Xi_4\}$
\end{minipage}\\
\noalign{\smallskip}\hline\noalign{\smallskip}
$A_{4,2}^a$ &
\begin{minipage}{2.7cm}
$[\Xi_1,\Xi_4]=a\Xi_1$\\
$[\Xi_2,\Xi_4]=\Xi_2$\\
$[\Xi_3,\Xi_4]=\Xi_2+\Xi_3$\\
$a\neq 0,1$
\end{minipage}&  
\begin{minipage}{3cm}
$\{\Xi_1\}$, $\{\Xi_2\}$, \\
$\{\Xi_3\}$, $\{\Xi_4\}$,\\
$\{\Xi_1+\alpha_1 \Xi_2\}$,\\
$\{\Xi_1+a_1 \Xi_3\}$
\end{minipage} &  
\begin{minipage}{3cm}
$\{\Xi_1,\Xi_2\}$, $\{\Xi_1,\Xi_3\}$,\\
$\{\Xi_1,\Xi_4\}$,$\{\Xi_2,\Xi_3\}$,\\
$\{\Xi_2,\Xi_4\}$,\\
$\{\Xi_1+a_1 \Xi_2,\Xi_3\}$,\\
$\{\Xi_1+\alpha_1 \Xi_3,\Xi_2\}$
\end{minipage}   &
\begin{minipage}{3cm}
$\{\Xi_1,\Xi_2,\Xi_3\}$, \\
$\{\Xi_1,\Xi_2,\Xi_4\}$,\\
$\{\Xi_2,\Xi_3,\Xi_4\}$
\end{minipage}\\
\noalign{\smallskip}\hline\noalign{\smallskip}
$A_{4,2}^1$ &
\begin{minipage}{2.7cm}
$[\Xi_1,\Xi_4]=\Xi_1$\\
$[\Xi_2,\Xi_4]=\Xi_2$\\
$[\Xi_3,\Xi_4]=\Xi_2+\Xi_3$
\end{minipage}&  
\begin{minipage}{3cm}
$\{\Xi_1\}$, $\{\Xi_2\}$, \\
$\{\Xi_3\}$, $\{\Xi_4\}$,\\
$\{\Xi_1+a_1 \Xi_2\}$,\\
$\{\Xi_1+a_1 \Xi_3\}$
\end{minipage} &  
\begin{minipage}{3cm}
$\{\Xi_1,\Xi_2\}$, $\{\Xi_1,\Xi_3\}$,\\
$\{\Xi_1,\Xi_4\}$, $\{\Xi_2,\Xi_3\}$,\\
$\{\Xi_2,\Xi_4\}$,\\
$\{\Xi_1+a_1 \Xi_2,\Xi_3\}$,\\
$\{\Xi_1+a_1 \Xi_2,\Xi_4\}$,\\
$\{\Xi_1+a_1 \Xi_3,\Xi_2\}$
\end{minipage}   &
\begin{minipage}{3cm}
$\{\Xi_1,\Xi_2,\Xi_3\}$, $\{\Xi_1,\Xi_2,\Xi_4\}$,\\
$\{\Xi_2,\Xi_3,\Xi_4\}$,\\
$\{\Xi_1+a_1 \Xi_3,\Xi_2,\Xi_4\}$
\end{minipage}\\
\noalign{\smallskip}\hline\noalign{\smallskip}
$A_{4,3}$ &
\begin{minipage}{2.7cm}
$[\Xi_1,\Xi_4]=\Xi_1$\\
$[\Xi_3,\Xi_4]=\Xi_2$
\end{minipage}&  
\begin{minipage}{3cm}
$\{\Xi_1\}$, $\{\Xi_2\}$, \\
$\{\Xi_3\}$, $\{\Xi_4\}$,\\
$\{\Xi_1+\alpha_1 \Xi_2\}$, \\
$\{\Xi_1+a_1 \Xi_3\}$,\\
$\{\Xi_3+a_1 \Xi_4\}$
\end{minipage} &  
\begin{minipage}{3cm}
$\{\Xi_1,\Xi_2\}$, $\{\Xi_1,\Xi_3\}$,\\
$\{\Xi_1,\Xi_4\}$, $\{\Xi_2,\Xi_3\}$,\\
$\{\Xi_2,\Xi_4\}$, \\
$\{\Xi_1,\Xi_3+a_1 \Xi_4\}$,\\
$\{\Xi_2,\Xi_3+a_1 \Xi_4\}$,\\
$\{\Xi_1+a_1 \Xi_2,\Xi_3\}$,\\
$\{\Xi_1+\alpha_1 \Xi_3,\Xi_2\}$
\end{minipage}   &
\begin{minipage}{3cm}
$\{\Xi_1,\Xi_2,\Xi_3\}$, $\{\Xi_1,\Xi_2,\Xi_4\}$,\\
$\{\Xi_2,\Xi_3,\Xi_4\}$,\\
$\{\Xi_1,\Xi_2,\Xi_3+a_1 \Xi_4\}$
\end{minipage}\\
\noalign{\smallskip}\hline\noalign{\smallskip}
$A_{4,4}$ &
\begin{minipage}{2.7cm}
$[\Xi_1,\Xi_4]=\Xi_1$\\
$[\Xi_2,\Xi_4]=\Xi_1+\Xi_2$\\
$[\Xi_3,\Xi_4]=\Xi_2+\Xi_3$ 
\end{minipage}&  
\begin{minipage}{3cm}
$\{\Xi_1\}$, $\{\Xi_2\}$, \\
$\{\Xi_3\}$, $\{\Xi_4\}$,\\
$\{\Xi_1+a_1 \Xi_3\}$
\end{minipage} &  
\begin{minipage}{3cm}
$\{\Xi_1,\Xi_2\}$, $\{\Xi_1,\Xi_3\}$,\\
$\{\Xi_1,\Xi_4\}$, $\{\Xi_2,\Xi_3\}$,\\
$\{\Xi_1+a_1 \Xi_3,\Xi_2\}$
\end{minipage}   &
\begin{minipage}{3cm}
$\{\Xi_1,\Xi_2,\Xi_3\}$, $\{\Xi_1,\Xi_2,\Xi_4\}$
\end{minipage}\\
\noalign{\smallskip}\hline\noalign{\smallskip}
$A_{4,5}^{a,b}$ &
\begin{minipage}{2.7cm}
$[\Xi_1,\Xi_4]=\Xi_1$\\
$[\Xi_2,\Xi_4]=a\Xi_2$\\
$[\Xi_3,\Xi_4]=b\Xi_3$\\
$-1\le a < b < 1$\\
$ab\neq 0$ 
\end{minipage}&  
\begin{minipage}{3cm}
$\{\Xi_1\}$, $\{\Xi_2\}$, \\
$\{\Xi_3\}$, $\{\Xi_4\}$,\\
$\{\Xi_1+\alpha_1 \Xi_2\}$, \\
$\{\Xi_1+\alpha_1 \Xi_3\}$,\\
$\{\Xi_2+\alpha_1 \Xi_3\}$,\\
$\{\Xi_1+\alpha_1 \Xi_2+a_1 \Xi_3\}$
\end{minipage} &  
\begin{minipage}{3cm}
$\{\Xi_1,\Xi_2\}$, $\{\Xi_1,\Xi_3\}$,\\
$\{\Xi_1,\Xi_4\}$, $\{\Xi_2,\Xi_3\}$,\\
$\{\Xi_2,\Xi_4\}$, $\{\Xi_3,\Xi_4\}$,\\
$\{\Xi_1,\Xi_2+\alpha_1 \Xi_3\}$,\\
$\{\Xi_1+\alpha_1 \Xi_2,\Xi_3\}$,\\
$\{\Xi_1+\alpha_1 \Xi_3,\Xi_2\}$, \\
$\{\Xi_1+\alpha_1 \Xi_3,\Xi_2+\alpha_2 \Xi_3\}$
\end{minipage}   &
\begin{minipage}{3cm}
$\{\Xi_1,\Xi_2,\Xi_3\}$, $\{\Xi_1,\Xi_2,\Xi_4\}$,\\
$\{\Xi_1,\Xi_3,\Xi_4\}$, $\{\Xi_2,\Xi_3,\Xi_4\}$
\end{minipage}\\
\noalign{\smallskip}\hline\noalign{\smallskip}
$A_{4,5}^{a,a}$ &
\begin{minipage}{2.7cm}
$[\Xi_1,\Xi_4]=\Xi_1$\\
$[\Xi_2,\Xi_4]=a\Xi_2$\\
$[\Xi_3,\Xi_4]=a\Xi_3$\\
$-1\le a < 1$ \\
$a\neq 0$
\end{minipage}&  
\begin{minipage}{3cm}
$\{\Xi_1\}$, $\{\Xi_2\}$, \\
$\{\Xi_3\}$, $\{\Xi_4\}$,\\
$\{\Xi_1+\alpha_1 \Xi_2\}$, \\
$\{\Xi_1+\alpha_1 \Xi_3\}$,\\
$\{\Xi_2+a_1 \Xi_3\}$,\\
$\{\Xi_1+\alpha_1 \Xi_2+a_1 \Xi_3\}$
\end{minipage} &  
\begin{minipage}{3cm}
$\{\Xi_1,\Xi_2\}$, $\{\Xi_1,\Xi_3\}$,\\
$\{\Xi_1,\Xi_4\}$, $\{\Xi_2,\Xi_3\}$,\\
$\{\Xi_2,\Xi_4\}$, $\{\Xi_3,\Xi_4\}$,\\
$\{\Xi_1,\Xi_2+a_1 \Xi_3\}$,\\
$\{\Xi_1+\alpha_1 \Xi_2,\Xi_3\}$,\\
$\{\Xi_1+\alpha_1 \Xi_3,\Xi_2\}$,\\
$\{\Xi_2+a_1 \Xi_3,\Xi_4\}$, \\
$\{\Xi_1+\alpha_1 \Xi_3,\Xi_2+a_1 \Xi_3\}$ 
\end{minipage}   &
\begin{minipage}{3cm}
$\{\Xi_1,\Xi_2,\Xi_3\}$, $\{\Xi_1,\Xi_2,\Xi_4\}$, \\
$\{\Xi_1,\Xi_3,\Xi_4\}$, $\{\Xi_2,\Xi_3,\Xi_4\}$, \\
$\{\Xi_1,\Xi_2+a_1 \Xi_3,\Xi_4\}$
\end{minipage}\\
\noalign{\smallskip}\hline\noalign{\smallskip}
$A_{4,5}^{a,1}$ &
\begin{minipage}{2.7cm}
$[\Xi_1,\Xi_4]=\Xi_1$\\
$[\Xi_2,\Xi_4]=a\Xi_2$\\
$[\Xi_3,\Xi_4]=\Xi_3$\\
$-1\le a < 1$\\
$a\neq 0$
\end{minipage}&  
\begin{minipage}{3cm}
$\{\Xi_1\}$, $\{\Xi_2\}$, \\
$\{\Xi_3\}$, $\{\Xi_4\}$, \\
$\{\Xi_1+\alpha_1 \Xi_2\}$, \\
$\{\Xi_1+a_1 \Xi_3\}$, \\
$\{\Xi_2+\alpha_1 \Xi_3\}$, \\
$\{\Xi_1+\alpha_1 \Xi_2+a_1 \Xi_3\}$
\end{minipage} &  
\begin{minipage}{3cm}
$\{\Xi_1,\Xi_2\}$, $\{\Xi_1,\Xi_3\}$, \\
$\{\Xi_1,\Xi_4\}$, $\{\Xi_2,\Xi_3\}$, \\
$\{\Xi_2,\Xi_4\}$, $\{\Xi_3,\Xi_4\}$, \\
$\{\Xi_1,\Xi_2+\alpha_1 \Xi_3\}$, \\
$\{\Xi_1+\alpha_1 \Xi_2,\Xi_3\}$, \\
$\{\Xi_1+a_1 \Xi_3,\Xi_2\}$, \\
$\{\Xi_1+a_1 \Xi_3,\Xi_4\}$, \\
$\{\Xi_1+a_1 \Xi_3,\Xi_2+\alpha_1 \Xi_3\}$
\end{minipage}   &
\begin{minipage}{3cm}
$\{\Xi_1,\Xi_2,\Xi_3\}$, $\{\Xi_1,\Xi_2,\Xi_4\}$,\\
$\{\Xi_1,\Xi_3,\Xi_4\}$, $\{\Xi_2,\Xi_3,\Xi_4\}$,\\
$\{\Xi_1+a_1 \Xi_3,\Xi_2,\Xi_4\}$
\end{minipage}\\
\noalign{\smallskip}\hline\noalign{\smallskip}
$A_{4,5}^{1,1}$ &
\begin{minipage}{2.7cm}
$[\Xi_1,\Xi_4]=\Xi_1$\\
$[\Xi_2,\Xi_4]=\Xi_2$\\
$[\Xi_3,\Xi_4]=\Xi_3$
\end{minipage}&  
\begin{minipage}{3cm}
$\{\Xi_1\}$, $\{\Xi_2\}$, \\
$\{\Xi_3\}$, $\{\Xi_4\}$,\\
$\{\Xi_1+a_1 \Xi_2\}$, \\
$\{\Xi_1+a_1 \Xi_3\}$, \\
$\{\Xi_2+a_1 \Xi_3\}$, \\
$\{\Xi_1+a_1 \Xi_2+a_2 \Xi_3\}$
\end{minipage} &  
\begin{minipage}{3cm}
$\{\Xi_1,\Xi_2\}$, $\{\Xi_1,\Xi_3\}$, \\
$\{\Xi_1,\Xi_4\}$, $\{\Xi_2,\Xi_3\}$, \\
$\{\Xi_2,\Xi_4\}$, $\{\Xi_3,\Xi_4\}$, \\
$\{\Xi_1,\Xi_2+a_1 \Xi_3\}$, \\
$\{\Xi_1+a_1 \Xi_2,\Xi_3\}$, \\
$\{\Xi_1+a_1 \Xi_2,\Xi_4\}$, \\
$\{\Xi_1+a_1 \Xi_3,\Xi_2\}$, \\
$\{\Xi_1+a_1 \Xi_3,\Xi_4\}$, \\
$\{\Xi_2+a_1 \Xi_3,\Xi_4\}$, \\
$\{\Xi_1+a_1 \Xi_3,\Xi_2+a_2 \Xi_3\}$,\\
$\{\Xi_1+a_1 \Xi_2+a_2 \Xi_3,\Xi_4\}$
\end{minipage}   &
\begin{minipage}{3cm}
$\{\Xi_1,\Xi_2,\Xi_3\}$, $\{\Xi_1,\Xi_2,\Xi_4\}$,\\
$\{\Xi_1,\Xi_3,\Xi_4\}$, $\{\Xi_2,\Xi_3,\Xi_4\}$,\\
$\{\Xi_1,\Xi_2+a_1 \Xi_3,\Xi_4\}$, \\
$\{\Xi_1+a_1 \Xi_2,\Xi_3,\Xi_4\}$, \\
$\{\Xi_1+a_1 \Xi_3,\Xi_2,\Xi_4\}$,\\
$\{\Xi_1+a_1 \Xi_3,\Xi_2+a_2 \Xi_3,$\\
$\phantom{\{}\Xi_4\}$
\end{minipage}\\
\noalign{\smallskip}\hline\noalign{\smallskip}
$A_{4,6}^{a,b}$ &
\begin{minipage}{2.7cm}
$[\Xi_1,\Xi_4]=a\Xi_1$\\
$[\Xi_2,\Xi_4]=b\Xi_2-\Xi_3$\\
$[\Xi_3,\Xi_4]=\Xi_2+b\Xi_3$\\
$a\neq 0, b\ge 0$
\end{minipage}&  
\begin{minipage}{3cm}
$\{\Xi_1\}$, $\{\Xi_2\}$, $\{\Xi_4\}$,\\
$\{\Xi_1+a_1 \Xi_2\}$
\end{minipage} &  
\begin{minipage}{3cm}
$\{\Xi_1,\Xi_2\}$, $\{\Xi_1,\Xi_4\}$, \\
$\{\Xi_2,\Xi_3\}$, \\
$\{\Xi_1+a_1 \Xi_2,\Xi_3\}$
\end{minipage}   &
\begin{minipage}{3cm}
$\{\Xi_1,\Xi_2,\Xi_3\}$, $\{\Xi_2,\Xi_3,\Xi_4\}$
\end{minipage}\\
\noalign{\smallskip}\hline\noalign{\smallskip}
$A_{4,7}$ &
\begin{minipage}{2.7cm}
$[\Xi_1,\Xi_4]=2\Xi_1$\\
$[\Xi_2,\Xi_4]=\Xi_2$\\
$[\Xi_3,\Xi_4]=\Xi_2+\Xi_3$\\
$[\Xi_2,\Xi_3]=\Xi_1$
\end{minipage}&  
\begin{minipage}{3cm}
$\{\Xi_1\}$, $\{\Xi_2\}$, \\
$\{\Xi_3\}$, $\{\Xi_4\}$
\end{minipage} &  
\begin{minipage}{3cm}
$\{\Xi_1,\Xi_2\}$, $\{\Xi_1,\Xi_3\}$,\\
$\{\Xi_1,\Xi_4\}$, $\{\Xi_2,\Xi_4\}$
\end{minipage}   &
\begin{minipage}{3cm}
$\{\Xi_1,\Xi_2,\Xi_3\}$, $\{\Xi_1,\Xi_2,\Xi_4\}$
\end{minipage}\\
\noalign{\smallskip}\hline\noalign{\smallskip}
$A_{4,8}$ &
\begin{minipage}{2.7cm}
$[\Xi_2,\Xi_3]=\Xi_1$\\
$[\Xi_2,\Xi_4]=\Xi_2$\\
$[\Xi_3,\Xi_4]=-\Xi_3$
\end{minipage}&  
\begin{minipage}{3cm}
$\{\Xi_1\}$, $\{\Xi_2\}$, \\
$\{\Xi_3\}$, $\{\Xi_4\}$,\\
$\{\Xi_2+\alpha_1 \Xi_3\}$, \\
$\{\Xi_1+a_1 \Xi_4\}$
\end{minipage} &  
\begin{minipage}{3cm}
$\{\Xi_1,\Xi_2\}$, $\{\Xi_1,\Xi_3\}$, \\
$\{\Xi_1,\Xi_4\}$, $\{\Xi_2,\Xi_4\}$,\\
$\{\Xi_3,\Xi_4\}$, \\
$\{\Xi_1,\Xi_2+\alpha_1 \Xi_3\}$,\\
$\{\Xi_1+a_1 \Xi_4,\Xi_2\}$,\\
$\{\Xi_1+a_1 \Xi_4,\Xi_3\}$
\end{minipage}   &
\begin{minipage}{3cm}
$\{\Xi_1,\Xi_2,\Xi_3\}$, $\{\Xi_1,\Xi_2,\Xi_4\}$,\\
$\{\Xi_1,\Xi_3,\Xi_4\}$
\end{minipage}\\
\noalign{\smallskip}\hline\noalign{\smallskip}
$A_{4,9}^{b}$ &
\begin{minipage}{2.7cm}
$[\Xi_2,\Xi_3]=\Xi_1$\\
$[\Xi_1,\Xi_4]=(1+b)\Xi_1$\\
$[\Xi_2,\Xi_4]=\Xi_2$\\
$[\Xi_3,\Xi_4]=b\Xi_3$\\
$0<|b|<1$
\end{minipage}&  
\begin{minipage}{3cm}
$\{\Xi_1\}$, $\{\Xi_2\}$, \\
$\{\Xi_3\}$, $\{\Xi_4\}$, \\
$\{\Xi_2+\alpha_1 \Xi_3\}$
\end{minipage} &  
\begin{minipage}{3cm}
$\{\Xi_1,\Xi_2\}$, $\{\Xi_1,\Xi_3\}$, \\
$\{\Xi_1,\Xi_4\}$, $\{\Xi_2,\Xi_4\}$, \\
$\{\Xi_3,\Xi_4\}$, \\
$\{\Xi_1,\Xi_2+\alpha_1 \Xi_3\}$
\end{minipage}   &
\begin{minipage}{3cm}
$\{\Xi_1,\Xi_2,\Xi_3\}$, $\{\Xi_1,\Xi_2,\Xi_4\}$,\\
$\{\Xi_1,\Xi_3,\Xi_4\}$
\end{minipage}\\
\noalign{\smallskip}\hline\noalign{\smallskip}
$A_{4,9}^{1}$ &
\begin{minipage}{2.7cm}
$[\Xi_2,\Xi_3]=\Xi_1$\\
$[\Xi_1,\Xi_4]=2\Xi_1$\\
$[\Xi_2,\Xi_4]=\Xi_2$\\
$[\Xi_3,\Xi_4]=\Xi_3$
\end{minipage}&  
\begin{minipage}{3cm}
$\{\Xi_1\}$, $\{\Xi_2\}$, \\
$\{\Xi_3\}$, $\{\Xi_4\}$,\\
$\{\Xi_2+a_1 \Xi_3\}$
\end{minipage} &  
\begin{minipage}{3cm}
$\{\Xi_1,\Xi_2\}$, $\{\Xi_1,\Xi_3\}$,\\
$\{\Xi_1,\Xi_4\}$, $\{\Xi_2,\Xi_4\}$, \\
$\{\Xi_3,\Xi_4\}$, \\
$\{\Xi_1,\Xi_2+a_1 \Xi_3\}$,\\
$\{\Xi_2+a_1 \Xi_3,\Xi_4\}$
\end{minipage}   &
\begin{minipage}{3cm}
$\{\Xi_1,\Xi_2,\Xi_3\}$, $\{\Xi_1,\Xi_2,\Xi_4\}$, \\
$\{\Xi_1,\Xi_3,\Xi_4\}$, \\
$\{\Xi_1,\Xi_2+a_1 \Xi_3,\Xi_4\}$
\end{minipage}\\
\noalign{\smallskip}\hline\noalign{\smallskip}
$A_{4,9}^{0}$ &
\begin{minipage}{2.7cm}
$[\Xi_2,\Xi_3]=\Xi_1$\\
$[\Xi_1,\Xi_4]=\Xi_1$\\
$[\Xi_2,\Xi_4]=\Xi_2$
\end{minipage}&  
\begin{minipage}{3cm}
$\{\Xi_1\}$, $\{\Xi_2\}$, \\
$\{\Xi_3\}$, $\{\Xi_4\}$, \\
$\{\Xi_2+\alpha_1 \Xi_3\}$,\\
$\{\Xi_3+a_1 \Xi_4\}$
\end{minipage} &  
\begin{minipage}{3cm}
$\{\Xi_1,\Xi_2\}$, $\{\Xi_1,\Xi_3\}$,\\
$\{\Xi_1,\Xi_4\}$, $\{\Xi_2,\Xi_4\}$,\\
$\{\Xi_3,\Xi_4\}$, \\
$\{\Xi_1,\Xi_2+\alpha_1 \Xi_3\}$,\\
$\{\Xi_1,\Xi_3+a_1 \Xi_4\}$
\end{minipage}   &
\begin{minipage}{3cm}
$\{\Xi_1,\Xi_2,\Xi_3\}$, $\{\Xi_1,\Xi_2,\Xi_4\}$,\\
$\{\Xi_1,\Xi_3,\Xi_4\}$, \\
$\{\Xi_1,\Xi_2,\Xi_3+a_1 \Xi_4\}$
\end{minipage}\\
\noalign{\smallskip}\hline\noalign{\smallskip}
$A_{4,10}$ &
\begin{minipage}{2.7cm}
$[\Xi_2,\Xi_3]=\Xi_1$\\
$[\Xi_2,\Xi_4]=-\Xi_3$\\
$[\Xi_3,\Xi_4]=\Xi_2$
\end{minipage}&  
\begin{minipage}{3cm}
$\{\Xi_1\}$, $\{\Xi_2\}$, $\{\Xi_4\}$,\\
$\{\Xi_1+a_1 \Xi_4\}$
\end{minipage} &  
\begin{minipage}{3cm}
$\{\Xi_1,\Xi_2\}$, $\{\Xi_1,\Xi_4\}$
\end{minipage}   &
\begin{minipage}{3cm}
$\{\Xi_1,\Xi_2,\Xi_3\}$
\end{minipage}\\
\noalign{\smallskip}\hline\noalign{\smallskip}
$A_{4,11}^{a}$ &
\begin{minipage}{2.7cm}
$[\Xi_2,\Xi_3]=\Xi_1$\\
$[\Xi_1,\Xi_4]=2a\Xi_1$\\
$[\Xi_2,\Xi_4]=a\Xi_2-\Xi_3$\\
$[\Xi_3,\Xi_4]=\Xi_2+a\Xi_3$\\
$a > 0$
\end{minipage}&  
\begin{minipage}{3cm}
$\{\Xi_1\}$, $\{\Xi_2\}$, $\{\Xi_4\}$
\end{minipage} &  
\begin{minipage}{3cm}
$\{\Xi_1,\Xi_2\}$, $\{\Xi_1,\Xi_4\}$
\end{minipage}   &
\begin{minipage}{3cm}
$\{\Xi_1,\Xi_2,\Xi_3\}$
\end{minipage}\\
\noalign{\smallskip}\hline\noalign{\smallskip}
$A_{4,12}$ &
\begin{minipage}{2.7cm}
$[\Xi_1,\Xi_3]=\Xi_1$\\
$[\Xi_2,\Xi_3]=\Xi_2$\\
$[\Xi_1,\Xi_4]=-\Xi_2$\\
$[\Xi_2,\Xi_4]=\Xi_1$
\end{minipage}&  
\begin{minipage}{3cm}
$\{\Xi_1\}$, $\{\Xi_3\}$, $\{\Xi_4\}$, \\
$\{\Xi_3+a_1 \Xi_4\}$
\end{minipage} &  
\begin{minipage}{3cm}
$\{\Xi_1,\Xi_2\}$, $\{\Xi_1,\Xi_3\}$, \\
$\{\Xi_3,\Xi_4\}$
\end{minipage}   &
\begin{minipage}{3cm}
$\{\Xi_1,\Xi_2,\Xi_3\}$, $\{\Xi_1,\Xi_2,\Xi_4\}$, \\
$\{\Xi_1,\Xi_2,\Xi_3+a_1 \Xi_4\}$
\end{minipage}\\
\noalign{\smallskip}\hline\noalign{\smallskip}
\end{longtable}
\end{center}
\normalsize

\section{Conclusions}
\label{sec:concl}

In this paper, the algorithmic construction and the analysis of the optimal systems have been performed using the package \symbolie running on \emph{Wolfram Mathematica}\texttrademark. 
The implementation of the algorithm relies on the definition of $p$-family of Lie subalgebras; this implies that the closure of a vector subspace
with respect to the Lie bracket can be ascertained without solving quadratic equations. We are aware that this approach may leave out some Lie subalgebras \cite{AmataOliveri2}, \emph{i.e.}, those Lie subalgebras not representing $p$-families. Anyway, in most of the cases this strategy is effective and provides the complete results. Then, introducing
a relation between families of subalgebras induced by the inner automorphisms (this relation is in general a preorder), we partition the families of Lie subalgebras and represent them by a graph: the simplest representatives of the connected
components of this graph give the list of optimal systems of families of Lie subalgebras. The results presented in this paper, dealing with all three- and four-dimensional Lie algebras, deeply investigated in \cite{PW}, represent a good test for the program. The results confirm those provided in \cite{PW}, with two exceptions. For algebra $A_{3,8}$, in \cite{PW} the one-dimensional optimal system contains the subalgebra $\{\Xi_1+\Xi_3\}$ which is not a family according to our definition, whereas \symbolie produces the family $\{\Xi_1+\alpha_1\Xi_3\}$, $\alpha_1=\pm 1$.
In the case of the algebra $A_{4,5}^{a,b}$, the results given in \cite{PW} miss the family $\{\Xi_1+\alpha_1\Xi_2\}$, that is listed by \symbolie. It is easy to recognize that this family must belong to the optimal system because it is invariant with respect to all the inner automorphisms of the Lie algebra. 

The analysis of higher dimensional Lie algebras is currently in progress, as well as the implementation of some strategies to speed up the algorithm. Anyway, we are confident that such a package in its present version can be useful to researchers.

\bigskip
\subsection*{Acknowledgments}
Work supported by ``Gruppo Nazionale per la Fisica Matematica'' of the ``Istituto Nazionale di Alta Matematica'', Italy.

\label{lastpage}

\end{document}